\documentclass[aps,prd,preprint,groupedaddress]{revtex4-2}

\usepackage{graphicx}
\usepackage{amsmath}
\usepackage{amsfonts}
\usepackage{amssymb}
\usepackage{slashed}
\usepackage{url}
\usepackage{hyperref}
\usepackage{subfigure}
\usepackage{float}
\usepackage{multirow}
\usepackage{array}
\usepackage{tikz}
\usepackage{tikz-feynman}
\tikzfeynmanset{compat=1.1.0}
\usepackage[force]{feynmp-auto}
\usepackage{feynmp}
\usepackage{microtype}

\allowdisplaybreaks

\begin{document}

\title{Probing the $\Lambda_{b}\to \Lambda_{c}^{*}\tau \bar{\nu}_{\tau}$ decays with leptoquarks}

\author{C P Haritha}
\email{harithacp2010@gmail.com}

\author{Karthik Jain}
\email{19phph08@uohyd.ac.in}

\author{Barilang Mawlong}
\email{barilang@uohyd.ac.in}

\affiliation{School of Physics,  University of Hyderabad, Hyderabad-500046,  India}

\begin{abstract}
 The extension of the standard model to include a single scalar or vector leptoquark has been shown to account for the observed deviations in the lepton flavor universality ratios in the $b \to c \ell \nu_{\ell}$ and $b \to s \ell \ell$ transitions. Exploring new physics in the $b \to c \tau \nu_{\tau}$ decays, in this work we analyze the baryonic decay channels $\Lambda_b \to \Lambda_{c}^{*}(2595,2625) \tau^- \bar{\nu}_\tau$ beyond the standard model. Specifically, we investigate the role of leptoquarks in these decays, focusing on the $U_1$ vector leptoquark and the $S_1$ scalar leptoquark. Employing recent lattice QCD results for the $\Lambda_b \to \Lambda_{c}^{*}$ form factors, the helicity amplitudes for all possible four-fermion interactions are worked out explicitly and presented here. Utilizing the current  $b \to c \ell \nu_{\ell}$ experimental data, we impose constraints on the leptoquark couplings and test new physics sensitivity of various observables for the decay processes under consideration. Our analysis demonstrates the significance of testing the lepton flavor universality ratio $R_{\Lambda_{c}^{*}}$ as it is observed to be particularly sensitive to these leptoquarks.
 \end{abstract}

\maketitle

\clearpage

\section{Introduction}
\label{sec:1}

The intriguing prospect of new physics beyond the standard model (SM) has lured significant attention, motivated by the persistent flavor anomalies observed in the $b$-sector and the abundance of data from current and anticipated forthcoming precision experiments. These anomalies offer indirect probes of new physics (NP) beyond the SM. The ratio of branching fractions that addresses lepton flavor universality (LFU), a key property inherent to the SM, has shown substantial tensions in its measurements with the SM in $b$-decay processes. In flavor-changing neutral current (FCNC) $b \to s \ell^+ \ell^-$ transitions, although the updated experimental measurements of the ratio $R_{K^{(*)}}=\mathcal{B}(B \to K^{(*)} \mu^+ \mu^-)/\mathcal{B}(B \to K^{(*)} e^+ e^-)$ are compatible with the SM predictions \cite{LHCb:2022vje}, there exists tensions at $2-4 \sigma$ level in other observables, such as $P_{5}^{\prime}$ of $B \to K^{*} \mu^{+} \mu^{-}$, the branching ratio and angular observables of $B_{s}\to \phi \mu^{+}\mu^{-}$ and $\mathcal{B}(B \to K \mu^{+} \mu^{-})$ \cite{Descotes-Genon:2012isb,LHCb:2020lmf,LHCb:2021zwz,LHCb:2014cxe}. Potential explanations for these discrepancies include modified couplings and the existence of new particles, such as leptoquarks (LQs) or additional gauge bosons like $Z^\prime$, predicted by various NP models.
In flavor-changing charged current (FCCC) $b \to c {\ell}^- \bar{\nu}_\ell$ transitions, the LFU ratios $R_{D^{(*)}}=\mathcal{B}(B \to D^{(*)}{\tau} \bar{\nu}_{\tau})/\mathcal{B}(B \to D^{(*)} \ell \bar{\nu}_{\ell}),$ with $\ell= e, \mu$, as reported by the HFLAV group \cite{HFLAV} continue to exhibit, but reduced, deviations. The measured values, $R_{D}^{expt}=0.342\pm 0.026$ and $R_{D^*}^{expt}=0.287\pm 0.012$, exceed the SM predictions by $1.6 \sigma$ and $2.5 \sigma$, respectively. The combined tension corresponds to about $3.31 \sigma$. In  $B_c \to J/\psi \ell \nu_\ell$ decay process, the combined average of the LHCb \cite{LHCb:2017vlu} and CMS \cite{CMS:2024seh,CMS:2024uyo} measurements for $R_{J/ \psi} = \mathcal{B} (B_c \to J/\psi \tau \bar{\nu}_{\tau)}/ \mathcal{B} (B_c \to J/\psi  \mu \bar{\nu}_{\mu})=0.61\pm 0.18$ \cite{Iguro:2024hyk} aligns with the SM prediction \cite{Harrison:2020nrv} within $1.9 \sigma$. Other experimental measurements available in this sector are the polarization of $\tau$, $P_{\tau}^{D^*}=-0.38\pm 0.51{}^{+0.21}_{-0.16}$ \cite{Belle:2016dyj, Belle:2017ilt} and $D^{*}$, $F_{L}^{D^*}=0.49\pm 0.05$ \cite{Belle:2019ewo, LHCb:2023ssl} in $B \to D^{*}{\tau} \bar{\nu}_{\tau}$ decay measured by the Belle collaboration. To explain the observed anomalies in charged current decays, several NP models featuring a $W^{\prime}$ boson, a charged Higgs boson, or LQs have been developed.

Leptoquarks are theoretical particles that can couple to both quarks and leptons simultaneously, making them strong contenders for explaining the anomalies observed in the $b \to c$ and $b \to s$ sectors. These color triplet bosons have nonzero baryon and lepton numbers and can have a spin of either $0$ (scalar) or $1$ (vector). They also possess a fractional electric charge. LQs inherently emerge in several extensions of the Standard Model, including technicolor models \cite{Farhi:1980xs, Lane:1991qh}, grand unified theories \cite{Georgi:1974sy, Fritzsch:1974nn}, Pati-Salam model \cite{Pati:1974yy}, $R$-parity-violating supersymmetric models \cite{Farrar:1978xj, Barbier:2004ez}  and quark-lepton composite models \cite{Schrempp:1984nj, Gripaios:2009dq}. 
On the experimental front, LQs have been thoroughly searched for via $e^{+}e^{-}, ep, pp, p\bar{p}, \gamma p$ and $e\gamma$ collisions \cite{Dorsner:2016wpm}. At the LHC, collaborations such as CMS and ATLAS  have dedicated programs for LQ searches \cite{CMS:2025lqs,ATLAS:2025gjo}. These searches have established stringent constraints on LQ masses and couplings. 
LQs as possible NP solutions for explaining the $b$-decay anomalies have been explored in literature \cite{Sakaki:2013bfa, Hiller:2014yaa, Bauer:2015knc,Bhattacharya:2016mcc, Dorsner:2016wpm,Dumont:2016xpj, Li:2016vvp,Becirevic:2016yqi, Das:2016vkr, Sahoo:2016pet, Chen:2017hir,Crivellin:2017zlb, Angelescu:2018tyl, Jung:2018lfu, Iguro:2018vqb, Popov:2019tyc, Crivellin:2019dwb, Aydemir:2019ynb, Iguro:2020keo, Cheung:2020sbq, Angelescu:2021lln,Alok:2017jaf,Alok:2017jgr}. In Ref. \cite{Angelescu:2021lln}, the authors demonstrated that the $U_1$ vector LQ model which transforms under the SM gauge group $SU(3)_C \times SU(2)_L \times U(1)_Y$ as $( \textbf{3}, \textbf{1}, 2/3 )$ can simultaneously address $R_{D^{(*)}}$ and $R_{K^{(*)}}$ anomalies. They also showed that the extension of the SM with a single scalar LQ such as $S_1 (\bar{\textbf{3}}$, $\textbf{1}, 1/3 )$ or $R_2 (\textbf{3}, \textbf{2}, 7/6)$ can account for $R_{D^{(*)}}$, while an $S_3(\bar{\textbf{3}}, \textbf{3}, 1/3 )$ scalar LQ or a $U_3(\textbf{3}, \textbf{3}, 2/3)$ vector LQ can accommodate the previously observed $R_{K^{(*)}}$ data. Thus, with LQs being favorable new physics candidates to explain flavor anomalies, in this paper, we study their effects on low-energy $b$-physics further. We particularly examine the semileptonic baryon $\Lambda_b \to \Lambda_c^{*} \ell \nu_\ell$ decays in the presence of the $U_1$ and $S_1$ LQs.
Since $b$-baryons have half-integer spin, their decays can provide access to a richer set of spin and angular observables, enhancing the sensitivity towards NP searches and providing additional avenues for testing the anomalies observed in $b$-meson decays.
Furthermore, semileptonic heavy baryon decays offer an effective framework for exploring heavy quark dynamics, with minimal contamination from non-perturbative QCD effects owing to the presence of leptons in the final state.

The $\Lambda_{b}^{0}$ baryon, which was also the first observed bottom baryon \cite{UA1:1991vse}, exhibits the simplest semileptonic $b$-baryon decay. The semileptonic decay of $\Lambda_{b}$ to the ground state $\Lambda_{c}$ with spin-parity quantum number, $J^{P}=\frac{1}{2}^{+}$  has been investigated within and beyond the SM considerably \cite{Ivanov:1996fj, Singleton:1990ye, Cheng:1995fe, Cardarelli:1998tq, Detmold:2015aaa, Gutsche:2015mxa, Faustov:2016pal, Zhao:2020mod, Li:2021qod, Duan:2022uzm, Zhang:2022bvl, Dutta:2015ueb, Shivashankara:2015cta, Li:2016pdv, Datta:2017aue, Ray:2018hrx, Bernlochner:2018bfn, Boer:2019zmp, Fedele:2022iib, Karmakar:2023rdt, Nandi:2024aia}. These $\Lambda_{b}$ decay modes offer a theoretically cleaner environment to examine the sub-leading order corrections owing to the strong constraints imposed by heavy quark symmetry \cite{Georgi:1990ei}. The DELPHI \cite{DELPHI:2003qft} and LHCb \cite{LHCb:2022piu} collaborations have measured the semileptonic $\Lambda_{b}$ decay modes. The shape of the differential decay rate for $\Lambda_b \to \Lambda_{c} \mu^- \bar{\nu_\mu}$ predicted by lattice QCD \cite{Detmold:2015aaa} was found to be in agreement with the LHCb measurement \cite{LHCb:2017vhq}. LHCb \cite{LHCb:2022piu} also measured the value of the LFU ratio, $R_{\Lambda_c}^{expt}=\mathcal{B}(\Lambda_b \to \Lambda_c \tau \bar{\nu}_{\tau})/\mathcal{B}(\Lambda_b \to \Lambda_c \mu \bar{\nu}_{\mu}) =0.242 \pm 0.026 \pm 0.071$, which is consistent with the SM prediction \cite{Bernlochner:2018kxh}. On examining backgrounds to the ground state decay $\Lambda_b \to \Lambda_c \mu \bar{\nu}_{\mu}$, dominant yields of $\Lambda_{c}^{*}(2595)$ and $\Lambda_{c}^{*}(2625)$ were observed \cite{LHCb:2017vhq}. This observation highlights the prospects for investigating LFU tests in the decays of these baryons as well. Here, the final hadron states $\Lambda_{c}^{*}(2595)$ and $\Lambda_{c}^{*}(2625)$ denotes the excited charmed baryons, with $J^{P}=\frac{1}{2}^{-}$ and $\frac{3}{2}^{-}$, respectively. These $\Lambda_{c}^{*}$ narrow resonances decay to $\Lambda_c \pi \pi$ states. CDF collaboration also measured the branching fractions of $\Lambda_{b}^{0} \to \Lambda_{c}(2595)^{+} \mu^- \bar{\nu_\mu}$ and $\Lambda_{b}^{0} \to \Lambda_{c}(2625)^{+} \mu^- \bar{\nu_\mu}$ relative to the branching fraction of $\Lambda_{b}^{0} \to \Lambda_{c}^{+} \mu^- \bar{\nu_\mu}$ \cite{CDF:2008hqh}. The current review of particle physics lists $\mathcal{B}(\Lambda_{b}^{0} \to \Lambda_{c}(2595)^{+} \mu^- \bar{\nu_\mu})=(7.9{}^{+4.0}_{-3.5})\times 10^{-3}$ and $\mathcal{B}(\Lambda_{b}^{0} \to \Lambda_{c}(2625)^{+} \mu^- \bar{\nu_\mu})=(1.3{}^{+0.6}_{-0.5})\%$ \cite{ParticleDataGroup:2024cfk}. 
The agreement of experimental results with SM predictions in the $\Lambda_b \to \Lambda_{c} \mu^- \bar{\nu_\mu}$ channel indicates that new physics prospects in $b \to c \mu \bar{\nu}_{\mu}$ transitions has miniscule scope. On the other hand, the tauonic modes remain desirable candidates for the discovery of NP signatures due to the continued discrepancies seen in $B \to D^{(*)} \tau \bar{\nu}_\tau$ decays. These channels are inherently more sensitive to potential NP contributions, such as scalar or tensor interactions that are helicity suppressed for lighter leptons. This encourages a careful exploration of LFU violation and the implications of NP in the $\tau$  modes of $\Lambda_b$ decays. Moreover, the presence of excited charmed baryon states such as $\Lambda_c^{*}(2595)$ and $\Lambda_c^{*}(2625)$ in semileptonic $\Lambda_b$ decays opens up an additional, relatively unexplored window to test LFU. Once reconstructed in the $\tau$ channel, these $\Lambda_b \to \Lambda_{c}^{*}\tau \bar{\nu}_\tau$ decays may provide complementary sensitivity to NP effects not visible in ground-state transitions. In \cite{Bernlochner:2021vlv}, it was projected that the uncertainty in measuring the LFU ratio $R_{\Lambda_{c}^{*}}$ at LHCb will significantly improve from Run 3 to Run 6, enhancing the potential to probe LFU violation and NP effects in $b \to c \tau \bar{\nu}_{\tau}$ transitions through these excited baryon channels.

In literature, the semileptonic decay of $\Lambda_b$ to the excited charmed baryon states of $\Lambda_c^{*}$ have been analyzed in the SM using various theoretical frameworks such as the constituent quark model \cite{Pervin:2005ve}, the covariant confined quark model \cite{Gutsche:2018nks} and lattice QCD (LQCD) \cite{Meinel:2021rbm, Meinel:2021mdj}. The $\Lambda_b \to \Lambda_{c}^{(*)} (1/2^{\pm})$ weak transition form factors were calculated using light-front quark model in \cite{Li:2021qod} and using light-cone sum rules in \cite{Duan:2024lnw}. In \cite{Becirevic:2020nmb}, they were calculated in the Bakamjian-Thomas relativistic framework combined with a spectroscopic model. Form factors for $\Lambda_b \to \Lambda_{c}^{*}  (3/2^{-})$ transition were computed in light-front quark model \cite{Li:2022hcn} and within light-cone sum rules \cite{Aliev:2023tpk}.
The $\Lambda_b \to \Lambda_{c}^{*}(1/2^{-},3/2^{-})$ decays were also studied within the heavy quark effective theory (HQET) framework. Using heavy quark expansion, form factors for these transitions were investigated up to $\mathcal{O}(1/m_{c,b})$ in \cite{Roberts:1992xm, Leibovich:1997az} and radiative corrections at $\mathcal{O}(\alpha_s)$ were calculated in \cite{Boer:2018vpx}. In \cite{Papucci:2021pmj}, the authors provided the HQET parametrization of form factors for these decay modes up to $\mathcal{O}(1/m_{c,b},\alpha_s)$ within and beyond the SM. They also presented preliminary predictions of $R_{\Lambda_{c}^{*}}$ within the SM and beyond. Further, they found tensions between the lattice results and HQET. To address this discrepancy, authors in \cite{DiRisi:2023npw} derived $1/{m_{c}^{2}}$ corrections to $\Lambda_b \to \Lambda_{c}^{*}(1/2^{-},3/2^{-})$ form factors within HQET. Implications for $\Lambda_b \to \Lambda_{c}^{*}(2595,2625) \ell^- \bar{\nu_\ell}$ decays using heavy quark spin symmetry were studied in \cite{Nieves:2019kdh, Du:2022rbf}. In \cite{Du:2022ipt}, the authors studied $\Lambda_b \to \Lambda_{c}^{*}(2595,2625) \tau^- \bar{\nu_\tau}$ decays using an effective theory approach and presented predictions for various decay observables obtained in SM and with different NP interactions. They calculated the decay observables using the tensor formalism \cite{Penalva:2021wye}  which provides an alternative approach to the helicity amplitude formalism employed in our analysis. Also, our work differs from \cite{Du:2022ipt} as we follow a model-dependent approach. In our analysis, we examine the $\Lambda_b \to \Lambda_{c}^{*}(2595,2625) \tau^- \bar{\nu_\tau}$ decay modes within the SM and in the $U_1$ and $S_1$ LQ models. 

The organisation of the paper is as follows. In Sec. \ref{sec:2}, we present the theoretical framework, including the LQ models and the LQ couplings to SM fermions which contribute to $b \to c \tau^- \bar{\nu_\tau}$ transitions. The helicity formalism employed to calculate the differential decay rate and other related observables, using the form factors obtained in the LQCD framework \cite{Meinel:2021mdj}, is also presented in this section. The calculated helicity amplitudes in the presence of NP for ${1/2}^{+} \to ({1/2}^{-},{3/2}^{-})$ transitions are presented explicitly here. In Sec. \ref{sec:3}, we present the fit analysis where the LQ couplings are constrained from the current experimental measurements of $R_{D^{(*)}}$, $R_{J/ \psi}$, $F_{L}^{D^*}$, $P_{\tau}^{D^*}$ and the upper bound of $\mathcal{B}(B_{c}^{+}\rightarrow \tau^{+}\nu_{\tau})$ using a $\chi^2$ fit. The $q^2$-spectra of the concerned observables of $\Lambda_b \to \Lambda_{c}^{*}(2595,2625) \tau^- \bar{\nu_\tau}$ decays are explored within the $U_1$ and $S_1$ LQ scenarios and the results are discussed here. We summarize and conclude our findings in Sec. \ref{sec:4}.

\section{Theoretical Model}
\label{sec:2}

\subsection{Effective Hamiltonian}
\label{sec:2-1}
Incorporating both SM and NP contributions, the effective Hamiltonian for decays facilitated by $b\to c l {\nu_l}$ quark-level transitions is expressed as \cite{Murgui:2019czp},
\begin{equation}
\mathcal{H}_{eff} = \frac{4G_F}{\sqrt{2}}V_{cb} [ \left( 1 + C_{V_L} \right) O_{V_L}+ C_{V_R} O_{V_R} + C_{S_R} O_{S_R} + C_{S_L} O_{S_L} + C_{T}O_{T}]+ {\rm{h.c.}} ,
\end{equation}
where $G_F$ is the Fermi constant and $V_{cb}$ is the CKM matrix element. The fermionic operators are given by 
\begin{eqnarray}
  &O_{V_{L,R}}&=\left( \bar{c}\gamma^\mu b_{L,R}\right) \left(\bar{\ell}_{L}\gamma_\mu \nu_{\ell_L} \right),~~~~~
 O_{S_{L,R}}=\left( \bar{c} b_{L,R}\right) \left(\bar{\ell_{R}} \nu_{\ell_L} \right), \nonumber \\
 &O_{T}&=\left( \bar{c}\sigma^{\mu \nu} b_{L}\right) \left(\bar{\ell_{R}}\sigma_{\mu \nu} \nu_{\ell_L} \right).
\end{eqnarray}
Their corresponding vector, scalar, and tensor Wilson coefficients are denoted by $C_{V_{L,R}}$, $C_{S_{L,R}}$, $C_T$. Here, we consider the neutrinos to be left-handed only.

\subsection{Leptoquark models} 
\label{sec:2-2}
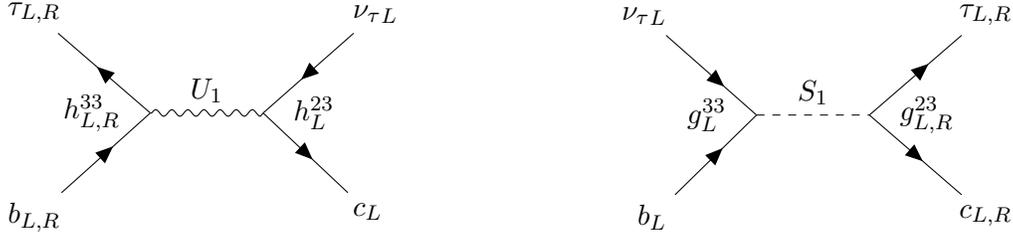
\begin{figure}[h]
\begin{subfigure}
\centering
\begin{tikzpicture}[baseline={(current bounding box.center)}]
\begin{feynman}
    \vertex (a) ;
    \vertex [above  left=of a] (c) {\(\tau_{L,R}\)};
    \vertex [below  left=of a] (d) {\(b_{L,R}\)};
    \vertex [      right=of a] (b) ;
    \vertex [above right=of b] (e) {\({\nu_\tau}_L\)};
    \vertex [below right=of b] (f) {\(c_L\)};
    \diagram* {
        (d) -- [fermion] (a) -- [fermion] (c),
        (a) -- [boson, edge label=\(U_1\)] (b),
        (e) -- [fermion] (b) -- [fermion] (f),
    };
    \vertex [left=0.7em of a] {\(h_{L,R}^{33}\)};
    \vertex [right=0.7em of b] {\(h_{L}^{23}\)};
\end{feynman}
\end{tikzpicture}
\end{subfigure}\hspace{0.15\textwidth}
\begin{subfigure}
\centering
\begin{tikzpicture}[baseline={(current bounding box.center)}]
\begin{feynman}
    \vertex (a) ;
    \vertex [above  left=of a] (c) {\({\nu_\tau}_L\)};
    \vertex [below  left=of a] (d) {\(b_L\)};
    \vertex [      right=of a] (b) ;
    \vertex [above right=of b] (e) {\(\tau_{L,R}\)};
    \vertex [below right=of b] (f) {\(c_{L,R}\)};
    \diagram* {
        (d) -- [fermion] (a) -- [anti fermion] (c),
        (a) -- [scalar, edge label=\(S_1\)] (b),
        (e) -- [anti fermion] (b) -- [fermion] (f),
    };
    \vertex [left=0.7em of a] {\(g_{L}^{33}\)};
    \vertex [right=0.7em of b] {\(g_{L,R}^{23}\)};
\end{feynman}
\end{tikzpicture}
\end{subfigure}
\caption{Feynman diagrams of $U_1$ and $S_1$ LQs mediating $b \to c \tau \bar{\nu}_{\tau}$ transition.}
\label{Feynman}
\end{figure}

The interaction between the SM fermions and the $U_1$ and $S_1$ LQs can be described, respectively, by \cite{Sakaki:2013bfa}
\begin{equation}\label{U1_lagrangian}
\mathcal{L}_{U_1}= h_{L}^{ij} \bar{Q}_{iL} \gamma_{\mu} L_{jL} U_{1}^{\mu} + h_{R}^{ij} \bar{d}_{iR} \gamma_{\mu}  \ell_{jR} U_{1}^{\mu} + {\rm{h.c.}},
\end{equation}
and
\begin{equation}\label{S1_lagrangian}
\mathcal{L}_{S_1}= g_{L}^{ij} \bar{Q}_{iL}^{c} i\sigma_2 L_{jL} S_{1}  + g_{R}^{ij} \bar{u}_{iR}^{c} \ell_{jR} S_{1} + {\rm{h.c.}},
\end{equation}
where $h_{L,R}^{ij}$ and $g_{L,R}^{ij}$ represent $3 \times 3$ complex matrices that characterize the couplings of $U_1$ and $S_1$ LQs with SM fermions, respectively. $Q_{iL}$ and $L_{jL}$ denote the SM left-handed quark and lepton doublets, while $u_{iR},d_{iR}$ and $\ell_{iR}$ represent the right-handed quark and lepton singlets, respectively. The indices $i,j$ refer to generation indices and $\psi^c = C \bar{\psi}^{T}= C \gamma^{0} \psi^{*} $ is a charge conjugated fermion field. The relevant Feynman diagrams showing the contribution of $U_1$ and $S_1$ LQs to $b \to c \tau \bar{\nu}_{\tau}$ transition are depicted in Fig. \ref{Feynman}.

On rotating the down-type quarks in Eq. (\ref{U1_lagrangian}) into the mass eigenstate basis and employing Fierz transformations, the resulting Wilson coefficients relevant to $b\to c \tau \bar{\nu}_{\tau}$ in the $U_1$ LQ model are found to be \cite{Sakaki:2013bfa} 
\begin{eqnarray}
C_{V_L} (\mu_{\mathrm{LQ}})&=& \frac{1}{2\sqrt{2}G_{F} V_{cb}}   \sum_{k=1}^{3} V_{k3} \frac{h_{L}^{23} {h_{L}^{k3*}}}{M_{U_1}^{2}} = \frac{1}{2\sqrt{2}G_{F} V_{cb}} V_{33} \frac{h_{L}^{23}{h_{L}^{33*}}}{M_{U_1}^{2}}, \label{CVL_U1}\\
C_{S_R} (\mu_{\mathrm{LQ}})&=& \frac{-1}{\sqrt{2}G_{F} V_{cb}} \sum_{k=1}^{3} V_{k3} \frac{h_{L}^{23}{h_{R}^{k3*}}}{M_{U_1}^{2}}=\frac{-1}{\sqrt{2}G_{F} V_{cb}} V_{33} \frac{h_{L}^{23}{h_{R}^{33*}}}{M_{U_1}^{2}}. \label{CSR_U1}
\end{eqnarray}
Similarly, applying the same procedure with Eq. (\ref{S1_lagrangian}) yields the new couplings in the $S_1$ LQ model as
\begin{eqnarray}
C_{V_L} (\mu_{\mathrm{LQ}})&=& \frac{1}{2\sqrt{2}G_{F} V_{cb}}   \sum_{k=1}^{3} V_{k3} \frac{g_{L}^{k3} {g_{L}^{23*}}}{2 M_{S_1}^{2}} = \frac{1}{2\sqrt{2}G_{F} V_{cb}} V_{33} \frac{g_{L}^{33} {g_{L}^{23*}}}{2 M_{S_1}^{2}} , \label{CVL_S1}\\
C_{S_L} (\mu_{\mathrm{LQ}})&=& \frac{-1}{2 \sqrt{2}G_{F} V_{cb}} \sum_{k=1}^{3} V_{k3} \frac{g_{L}^{k3} {g_{R}^{23*}}}{2 M_{S_1}^{2}} =\frac{-1}{2 \sqrt{2}G_{F} V_{cb}} V_{33} \frac{g_{L}^{33} {g_{R}^{23*}}}{2 M_{S_1}^{2}}, \label{CSL_S1}\\
C_{T} (\mu_{\mathrm{LQ}})&=& \frac{1}{2 \sqrt{2}G_{F} V_{cb}} \sum_{k=1}^{3} V_{k3} \frac{g_{L}^{k3} {g_{R}^{23*}}}{8 M_{S_1}^{2}} =\frac{1}{2 \sqrt{2}G_{F} V_{cb}} V_{33} \frac{g_{L}^{33} {g_{R}^{23*}}}{8 M_{S_1}^{2}}, \label{CT_S1}
\end{eqnarray}
where $V_{k3}$ represents the CKM matrix elements, $M_{U_1}$ and $M_{S_1}$ indicates the masses of the corresponding LQs. The pair-production searches for LQs at the collider experiments exclude LQ masses below $1.5$ TeV \cite{CMS:2018qqq, CMS:2018iye, CMS:2022nty, ATLAS:2023kek}. In our calculation, we adopt a benchmark LQ mass of $2$ TeV. In deriving the final expressions, we omit the subdominant CKM elements $V_{13}$ and $V_{23}$.

The Wilson coefficients presented in Eqs. (\ref{CVL_U1})-(\ref{CT_S1}) are defined at the LQ scale. However the relevant hadronic processes occur at the $m_b$ scale.  In our analysis, we have employed the renormalization group evolution (RGE) equations to evolve the scalar and tensor coefficients to the $m_b$ scale. Due to vector current conservation, $C_{V_L}$ is not renormalized \cite{Li:2016pdv}. The RGE equations corresponding to the LQ scale of $2$ TeV can be found in \cite{Iguro:2024hyk}.

\subsection{Hadronic matrix elements}
\label{sec:2-3}
The matrix elements of the different transition operators between the baryon states for $B_{1} \to B_{2}^{*} \ell^{-} \bar{\nu}_{\ell}$ (where $B_1 = \Lambda_b, B_2^* = \Lambda_c^*)$ decays can be parametrized in terms of various form factors.  For the ${1/2}^{+} \to {1/2}^{-}$ transition, the hadronic matrix elements for the vector and axial-vector currents are given by \cite{Du:2022ipt}
\begin{eqnarray}
\left\langle B_2^*  \vert \bar{c}\gamma^{\mu}b \vert B_1 \right\rangle &=& \bar{u}_{B_2^*}(p_2,s_2)\Big[ G_1 \gamma^{\mu} + G_2 \frac{p_{1}^{\mu}}{M_1} + G_3 \frac{p_{2}^{\mu}}{M_2} \Big]\gamma_5 u_{B_1}(p_1,s_1), \nonumber \\
\left\langle B_2^*  \vert \bar{c}\gamma^{\mu} \gamma_5 b \vert B_1 \right\rangle &=& \bar{u}_{B_2^*}(p_2,s_2)\Big[ F_1 \gamma^{\mu} + F_2 \frac{p_{1}^{\mu}}{M_1} + F_3 \frac{p_{2}^{\mu}}{M_2} \Big] u_{B_1}(p_1,s_1),
\label{eqn:MVA1}
\end{eqnarray}
where $p_1$ and $p_2$ represent the four-momenta, while $M_1$ and $M_2$ denote the masses of the parent baryon $B_1$ and daughter baryon $B_{2}^{*}$, respectively. The quantities $G_i$ and $F_i$ are the relevant form factors. $u_{B_1}$ and $\bar{u}_{B_2^*}$ are the Dirac spinors associated with $B_1$ and $B_{2}^{*}$, respectively.
 
Using Dirac equations, the hadronic matrix elements of (pseudo)scalar currents are extracted from Eq. (\ref{eqn:MVA1}) as
\begin{eqnarray}
\langle B_2^* \vert \bar{c}b \vert B_1 \rangle &=&\frac{q_\mu}{m_b-m_c} \langle B_2^* \vert \bar{c}\gamma^{\mu}b \vert B_1 \rangle \nonumber \\ 
&=& \bar{u}_{B_2^*}(p_2,s_2) \Big[ G_{1} \frac{\slashed{q} }{(m_b-m_c)} + G_{2}  \left( \frac{M_{1}^{2}-M_{2}^{2}+q^{2}}{2M_{1}(m_b-m_c)}\right) \nonumber \\ 
&&+ G_{3} \left( \frac{M_{1}^{2}-M_{2}^{2}-q^{2}}{2M_{2}(m_b-m_c)}\right) \Big] \gamma_5    u_{B_1} (p_1,s_1), \nonumber  \\
\langle B_2^* \vert \bar{c} \gamma_{5} b \vert B_1 \rangle &=& - \frac{q_\mu}{(m_b+m_c)}  \langle B_2^* \vert \bar{c}\gamma^{\mu}\gamma_{5}b   \vert B_1 \rangle  \nonumber \\ 
&=& \bar{u}_{B_2^*}(p_2,s_2) \Big[ -F_{1} \frac{\slashed{q} }{(m_b+m_c)} - F_{2}  \left( \frac{M_{1}^{2}-M_{2}^{2}+q^{2}}{2M_{1}(m_b+m_c)}\right) \nonumber \\ 
&&- F_{3} \left( \frac{M_{1}^{2}-M_{2}^{2}-q^{2}}{2M_{2}(m_b+m_c)}\right) \Big]   u_{B_1} (p_1,s_1),
\label{eqn:MSP1}
\end{eqnarray} 
where $q=p_1 - p_2$ is the four-momentum transfer, $m_b$ and $m_c$ are the masses of $b$ and $c$ quarks, respectively, evaluated at the scale $\mu=m_b$.

The tensor and pseudo-tensor hadronic matrix elements are parametrized as \cite{Du:2022ipt} 
\begin{eqnarray}
\langle B_2^* \vert \bar{c} i \sigma^{\mu \nu} b \vert B_1 \rangle &=&\bar{u}_{B_2^*}(p_2,s_2) \epsilon^{\mu \nu}{}_{\alpha \beta} \Big[ i \frac{T_1}{M_{1}^{2}} p_{1}^{\alpha} p_{2}^{\beta} + i \frac{T_2}{M_{1}} \gamma^{\alpha} p_{1}^{\beta} + i \frac{T_3}{M_{1}} \gamma^{\alpha} p_{2}^{\beta} \nonumber\\ &&+ i \frac{T_4}{2} \gamma^{\alpha} \gamma^{\beta} \Big] u_{B_1} (p_1,s_1), \nonumber \\
\langle B_2^* \vert \bar{c} i \sigma^{\mu \nu} \gamma_5 b \vert B_1 \rangle &=&\bar{u}_{B_2^*}(p_2,s_2)  \Big[ -\frac{T_1}{M_{1}^{2}} ( p_{1}^{\mu} p_{2}^{\nu} - p_{1}^{\nu} p_{2}^{\mu} ) - \frac{T_2}{M_{1}} ( \gamma^{\mu} p_{1}^{\nu} -\gamma^{\nu} p_{1}^{\mu} ) \nonumber\\ 
&&- \frac{T_3}{M_{1}} ( \gamma^{\mu} p_{2}^{\nu} - \gamma^{\nu} p_{2}^{\mu} ) + i T_4 \sigma^{\mu \nu} \Big] u_{B_1} (p_1,s_1). 
\label{eqn:MT1}
\end{eqnarray}
In the above, $\sigma^{\mu \nu}=\frac{i}{2}(\gamma^\mu \gamma^\nu - \gamma^\nu \gamma^\mu)$.

The form factors $G_i. F_i, T_i$ defined in Eqs. (\ref{eqn:MVA1})-(\ref{eqn:MT1}) are related to the helicity form factors computed in the LQCD framework \cite{Meinel:2021rbm} as
\begin{eqnarray}
G_1&=& -f_{\perp}^{(\frac{1}{2}^{-})}, \nonumber \\
G_2&=& M_1 \Big[ f_{0}^{(\frac{1}{2}^{-})} \frac{M_1+M_2}{q^2} +f_{+}^{(\frac{1}{2}^{-})} \frac{M_1-M_2}{Q_-} \left( 1- \frac{M_1^2-M_2^2}{q^2} \right) + f_{\perp}^{(\frac{1}{2}^{-})} \frac{2 M_2}{Q_-} \Big], \nonumber \\
G_3&=& M_2 \Big[ -f_{0}^{(\frac{1}{2}^{-})} \frac{M_1+M_2}{q^2} +f_{+}^{(\frac{1}{2}^{-})} \frac{M_1-M_2}{Q_-} \left( 1+ \frac{M_1^2-M_2^2}{q^2} \right) - f_{\perp}^{(\frac{1}{2}^{-})} \frac{2 M_1}{Q_-} \Big], \nonumber \\
F_1&=& -g_{\perp}^{(\frac{1}{2}^{-})}, \nonumber \\
F_2&=& M_1 \Big[ -g_{0}^{(\frac{1}{2}^{-})} \frac{M_1-M_2}{q^2} -g_{+}^{(\frac{1}{2}^{-})} \frac{M_1+M_2}{Q_+} \left( 1- \frac{M_1^2-M_2^2}{q^2} \right) + g_{\perp}^{(\frac{1}{2}^{-})} \frac{2 M_2}{Q_+} \Big], \nonumber \\
F_3&=& M_2 \Big[ g_{0}^{(\frac{1}{2}^{-})} \frac{M_1-M_2}{q^2} -g_{+}^{(\frac{1}{2}^{-})} \frac{M_1+M_2}{Q_+} \left( 1+ \frac{M_1^2-M_2^2}{q^2} \right) + g_{\perp}^{(\frac{1}{2}^{-})} \frac{2 M_1}{Q_+} \Big],  \nonumber\\
T_1&=& \frac{2 M_1^2}{Q_+} \Big[ h_{+}^{(\frac{1}{2}^{-})} - \tilde{h}_{+}^{(\frac{1}{2}^{-})} - \frac{Q_+ (M_1-M_2)^2}{q^2 Q_-} ( h_{\perp}^{(\frac{1}{2}^{-})} - h_{+}^{(\frac{1}{2}^{-})} ) \nonumber \\
&&+ \frac{(M_1+M_2)^2}{q^2} ( \tilde{h}_{\perp}^{(\frac{1}{2}^{-})} - h_{+}^{(\frac{1}{2}^{-})}  ) \Big], \nonumber \\
T_2&=& -\frac{2 M q.p_2}{q^2 Q_-} ( h_{\perp}^{(\frac{1}{2}^{-})} - h_{+}^{(\frac{1}{2}^{-})} ) (M_1 - M_2) + \frac{M_1}{q^2} ( \tilde{h}_{\perp}^{(\frac{1}{2}^{-})} - h_{+}^{(\frac{1}{2}^{-})} ) (M_1+M_2), \nonumber \\
T_3&=& \frac{2 M q.p_1}{q^2 Q_-} ( h_{\perp}^{(\frac{1}{2}^{-})} - h_{+}^{(\frac{1}{2}^{-})} ) (M_1 - M_2) - \frac{M_1}{q^2} ( \tilde{h}_{\perp}^{(\frac{1}{2}^{-})} - h_{+}^{(\frac{1}{2}^{-})} ) (M_1+M_2), \nonumber \\
T_4&=&  h_{+}^{(\frac{1}{2}^{-})},
\label{eqn:LQCDff1}
\end{eqnarray}
where $Q_{\pm}= (M_1 \pm M_2)^2-q^2$.

Likewise, for the ${1/2}^{+} \to {3/2}^{-}$ transition, the hadronic matrix elements for the (axial)vector, (pseudo)scalar and (pseudo)tensor interactions are parametrized as \cite{Du:2022ipt}
\begin{eqnarray}
	\left\langle B_2^*  \vert \bar{c}\gamma^{\mu}b \vert B_1 \right\rangle &=& \bar{u}_{B_2^*}^{\lambda}(p_2,s_2)\Big[ \frac{F_{1}^{V}}{M_1} p_{1 \lambda} \gamma^{\mu} + \frac{F_{2}^{V}}{M_1^2} p_{1\lambda} p_{1}^{\mu} + \frac{F_3^V}{M_1 M_2} p_{1\lambda} p_{2}^{\mu} + F_{4}^{V} g_{\lambda}{}^{\mu} \Big] u_{B_1}(p_1,s_1), \nonumber \\
	\left\langle B_2^*  \vert \bar{c}\gamma^{\mu} \gamma_5 b \vert B_1 \right\rangle &=&  \bar{u}_{B_2^*}^{\lambda}(p_2,s_2)\Big[ \frac{F_{1}^{A}}{M_1} p_{1 \lambda} \gamma^{\mu} + \frac{F_{2}^{A}}{M_1^2} p_{1\lambda} p_{1}^{\mu} + \frac{F_3^A}{M_1 M_2} p_{1\lambda} p_{2}^{\mu} + F_{4}^{A} g_{\lambda}{}^{\mu} \Big] \gamma_5 u_{B_1}(p_1,s_1), \nonumber \\
	\langle B_2^* \vert \bar{c}b \vert B_1 \rangle &=& \bar{u}_{B_2^*}^{\lambda}(p_2,s_2) \Big[ F_{1}^{V} p_{1 \lambda}  \frac{\slashed{q} }{M_1(m_b-m_c)} + F_{2}^{V}p_{1\lambda}  \left( \frac{M_{1}^{2}-M_{2}^{2}+q^{2}}{2M_{1}^2(m_b-m_c)}\right) \nonumber \\ 
	&&+ F_{3}^{V}p_{1\lambda} \left( \frac{M_{1}^{2}-M_{2}^{2}-q^{2}}{2M_1 M_{2}(m_b-m_c)}\right) + F_4^V \frac{q_\lambda}{(m_b-m_c)} \Big]   u_{B_1} (p_1,s_1), \nonumber  \\
	\langle B_2^* \vert \bar{c} \gamma_{5} b \vert B_1 \rangle &=& \bar{u}_{B_2^*}^{\lambda}(p_2,s_2) \Big[- F_{1}^{A} p_{1 \lambda}  \frac{\slashed{q} }{M_1(m_b+m_c)} - F_{2}^{A}p_{1\lambda}  \left( \frac{M_{1}^{2}-M_{2}^{2}+q^{2}}{2M_{1}^2(m_b+m_c)}\right) \nonumber \\ 
	&&- F_{3}^{A}p_{1\lambda} \left( \frac{M_{1}^{2}-M_{2}^{2}-q^{2}}{2M_1 M_{2}(m_b+m_c)}\right) - F_4^A \frac{q_\lambda}{(m_b+m_c)} \Big]  \gamma_5   u_{B_1} (p_1,s_1), \nonumber \\
	\langle B_2^* \vert \bar{c} i \sigma^{\mu \nu} b \vert B_1 \rangle &=&\bar{u}_{B_2^*}^{\lambda}(p_2,s_2)  \Big[ -\frac{F_1^T}{M_{1}^{3}} p_{1 \lambda} ( p_{1}^{\mu} p_{2}^{\nu} - p_{1}^{\nu} p_{2}^{\mu} ) - \frac{F_2^T}{M_{1}^2} p_{1 \lambda}( \gamma^{\mu} p_{1}^{\nu} -\gamma^{\nu} p_{1}^{\mu} ) \nonumber\\ 
	&&- \frac{F_3^T}{M_{1}^2} p_{1 \lambda} ( \gamma^{\mu} p_{2}^{\nu} - \gamma^{\nu} p_{2}^{\mu} ) + i \frac{F_4^T}{M_1} p_{1 \lambda} \sigma^{\mu \nu} - F_{5}^{T}( g_{\lambda}{}^{\mu} \gamma^{\nu}-g_{\lambda}{}^{\nu} \gamma^{\mu} ) \nonumber \\
	&&- \frac{F_6^T}{M_1} ( g_{\lambda}{}^{\mu} p_{1}^{\nu}- g_{\lambda}{}^{\nu} p_{1}^{\mu} ) - \frac{F_7^T}{M_1} ( g_{\lambda}{}^{\mu} p_{2}^{\nu}- g_{\lambda}{}^{\nu} p_{2}^{\mu} ) \Big] u_{B_1} (p_1,s_1), \nonumber \\
	\langle B_2^* \vert \bar{c} i \sigma^{\mu \nu} \gamma_5 b \vert B_1 \rangle &=&\bar{u}_{B_2^*}^{\lambda}(p_2,s_2) \gamma_5 \Big[ p_{1 \lambda} ( p_{1}^{\mu} p_{2}^{\nu} - p_{1}^{\nu} p_{2}^{\mu} ) \frac{F_1^T}{M_{1}^{3}} -  p_{1 \lambda}( \gamma^{\mu} p_{1}^{\nu} -\gamma^{\nu} p_{1}^{\mu} )  \nonumber\\ 
	&&\times \frac{M_2 F_1^T - M_1 F_2^T}{M_{1}^3} -  p_{1 \lambda} ( \gamma^{\mu} p_{2}^{\nu} - \gamma^{\nu} p_{2}^{\mu} )\frac{F_1^T + F_3^T}{M_{1}^2} \nonumber \\
	&&+ i p_{1 \lambda} \sigma^{\mu \nu} \Big[ \frac{F_6^T}{M_1}-\frac{F_1^T}{M_1^3}(p.p^{\prime}+ M_1 M_2)+\frac{F_2^T}{M_1}-\frac{M_2}{M_1^2} F_3^T +\frac{F_4^T}{M_1} \Big] \nonumber \\
	&&- ( g_{\lambda}{}^{\mu} \gamma^{\nu}-g_{\lambda}{}^{\nu} \gamma^{\mu} )(F_{5}^{T}+ F_{6}^{T} +\frac{M_2}{M_1} F_{6}^{T}) \nonumber \\
	&&+ ( g_{\lambda}{}^{\mu} p_{1}^{\nu}- g_{\lambda}{}^{\nu} p_{1}^{\mu} )\frac{F_6^T}{M_1}  -  ( g_{\lambda}{}^{\mu} p_{2}^{\nu}- g_{\lambda}{}^{\nu} p_{2}^{\mu} )\frac{F_7^T}{M_1} \Big] u_{B_1} (p_1,s_1). 
	\label{eqn:MT2}
\end{eqnarray}
In the above, $\bar{u}_{B_2^*}^{\lambda}$ denotes the Rarita-Schwinger spinor of $\Lambda_{c}^{*}(3/2^-)$ baryon.

The relations between the form factors defined in Eq. (\ref{eqn:MT2}) and the ones defined in \cite{Meinel:2021rbm} are given as \cite{Du:2022ipt}
\begin{eqnarray}
F_1^V&=& (f_{\perp}^{(\frac{3}{2}^{-})}+f_{\perp^\prime}^{(\frac{3}{2}^{-})})\frac{M_1 M_2}{Q_-}, \nonumber \\
F_2^V&=& M_1^2 \Big[ f_{0}^{(\frac{3}{2}^{-})} \frac{M_2}{Q_+}\frac{(M_1-M_2)}{q^2}+ f_{+}^{(\frac{3}{2}^{-})}\frac{M_2}{Q_-}\frac{(M_1+M_2)[q^2-(M_1^2-M_2^2)]}{q^2 Q_{+}}\nonumber\\
&&- (f_{\perp}^{(\frac{3}{2}^{-})}-f_{\perp^\prime}^{(\frac{3}{2}^{-})})\frac{2 M_2^2}{Q_- Q_+} \Big], \nonumber\\
F_3^V&=& M_2^2 \Big[ -f_{0}^{(\frac{3}{2}^{-})} \frac{M_1}{Q_+}\frac{(M_1-M_2)}{q^2}+ f_{+}^{(\frac{3}{2}^{-})}\frac{M_1}{Q_-}\frac{(M_1+M_2)[q^2+(M_1^2-M_2^2)]}{q^2 Q_{+}}\nonumber\\
&&- \Big[ f_{\perp}^{(\frac{3}{2}^{-})}-f_{\perp^\prime}^{(\frac{3}{2}^{-})}\left( 1- \frac{Q_+}{M_1 M_2} \right)  \Big] \frac{2 M_1^2}{Q_- Q_+} \Big], \nonumber\\
F_4^V&=&f_{\perp^\prime}^{(\frac{3}{2}^{-})}, \nonumber \\
F_1^A&=& (g_{\perp}^{(\frac{3}{2}^{-})}+g_{\perp^\prime}^{(\frac{3}{2}^{-})})\frac{M_1 M_2}{Q_+}, \nonumber \\
F_2^A&=& M_1^2 \Big[ -g_{0}^{(\frac{3}{2}^{-})} \frac{M_2}{Q_-}\frac{(M_1+M_2)}{q^2}-g_{+}^{(\frac{3}{2}^{-})}\frac{M_2}{Q_+}\frac{(M_1-M_2)[q^2-(M_1^2-M_2^2)]}{q^2 Q_{-}}\nonumber\\
&&- (g_{\perp}^{(\frac{3}{2}^{-})}-g_{\perp^\prime}^{(\frac{3}{2}^{-})})\frac{2 M_2^2}{Q_- Q_+} \Big], \nonumber\\
F_3^A&=& M_2^2 \Big[ g_{0}^{(\frac{3}{2}^{-})} \frac{M_1}{Q_-}\frac{(M_1+M_2)}{q^2}- g_{+}^{(\frac{3}{2}^{-})}\frac{M_1}{Q_+}\frac{(M_1+M_2)[q^2+(M_1^2-M_2^2)]}{q^2 Q_{-}}\nonumber\\
&&+ \Big[ g_{\perp}^{(\frac{3}{2}^{-})}-g_{\perp^\prime}^{(\frac{3}{2}^{-})}\left( 1+ \frac{Q_-}{M_1 M_2} \right)  \Big] \frac{2 M_1^2}{Q_- Q_+} \Big], \nonumber\\
F_4^A&=&g_{\perp^\prime}^{(\frac{3}{2}^{-})}, \nonumber\\
\label{eqn:LQCDffVA2}
F_1^T&=& -\frac{2 M_1^3 M_2}{Q_+ Q_-} (h_{+}^{(\frac{3}{2}^{-})} - \tilde{h}_{+}^{(\frac{3}{2}^{-})} ) - \frac{2 M_1^3 M_2 (M_1-M_2)^2 }{Q_+ Q_- q^2} \tilde{h}_{\perp}^{(\frac{3}{2}^{-})} \nonumber \\ 
&&+ \frac{2 M_1^3 (M_1-M_2)(M_1^2-M_1 M_2-q^2)}{Q_+ Q_- q^2} \tilde{h}_{\perp^\prime}^{(\frac{3}{2}^{-})} + \frac{2 M_1^3 M_2 (M_1+M_2)^2 }{Q_+ Q_- q^2} {h}_{\perp}^{(\frac{3}{2}^{-})} \nonumber \\ 
&&+ \frac{2 M_1^3 (M_1+M_2)(M_1^2+M_1 M_2-q^2)}{Q_+ Q_- q^2} {h}_{\perp^\prime}^{(\frac{3}{2}^{-})} ,\nonumber \\
F_2^T&=& \frac{2 M_1^2 M_2^2}{Q_+ Q_-} \tilde{h}_{+}^{(\frac{3}{2}^{-})} - \frac{M_1^2 M_2 (M_1-M_2)(M_1^2-M_2^2-q^2)}{Q_+ Q_- q^2} (\tilde{h}_{\perp}^{(\frac{3}{2}^{-})} - \tilde{h}_{\perp^\prime}^{(\frac{3}{2}^{-})}) \nonumber \\
&&+\frac{M_1^2 M_2 (M_1+M_2)}{Q_- q^2} ({h}_{\perp}^{(\frac{3}{2}^{-})} + {h}_{\perp^\prime}^{(\frac{3}{2}^{-})}), \nonumber \\
F_3^T&=& -\frac{2 M_1^3 M_2}{Q_+ Q_-} \tilde{h}_{+}^{(\frac{3}{2}^{-})} + \frac{M_1^2 M_2 (M_1-M_2)(M_1^2-M_2^2+q^2)}{Q_+ Q_- q^2} \tilde{h}_{\perp}^{(\frac{3}{2}^{-})}  \nonumber \\
&&- \frac{M_1 (M_1-M_2) (M_1^2+M_2^2-M_1 M_2-q^2) (M_1^2-M_2^2+q^2)}{Q_+ Q_- q^2} \tilde{h}_{\perp^\prime}^{(\frac{3}{2}^{-})} \nonumber \\
&&-\frac{M_1^2 M_2 (M_1+M_2)}{Q_- q^2} {h}_{\perp}^{(\frac{3}{2}^{-})} - \frac{M_1 (M_1^3 + M_2^3 - q^2(M_1+M_2))}{Q_- q^2} {h}_{\perp^\prime}^{(\frac{3}{2}^{-})}, \nonumber \\
F_4^T&=& \frac{M_1 M_2}{Q_+} \tilde{h}_{+}^{(\frac{3}{2}^{-})} -  \frac{M_2 (M_1+M_2)}{q^2} {h}_{\perp^\prime}^{(\frac{3}{2}^{-})}- \frac{M_2 (M_1-M_2)(M_1^2-M_2^2+q^2)}{Q_+ q^2} \tilde{h}_{\perp^\prime}^{(\frac{3}{2}^{-})},  \nonumber \\
F_5^T&=& -\frac{1}{2 M_1 q^2} \Big[ {h}_{\perp^\prime}^{(\frac{3}{2}^{-})} (M_1+M_2) Q_+ +\tilde{h}_{\perp^\prime}^{(\frac{3}{2}^{-})} (M_1-M_2)(M_1^2 -M_2^2 +q^2)\Big], \nonumber\\
F_6^T&=& \frac{1}{q^2} \Big[ {h}_{\perp^\prime}^{(\frac{3}{2}^{-})} (M_1+M_2)^{2} +\tilde{h}_{\perp^\prime}^{(\frac{3}{2}^{-})} (M_1-M_2)^{2} \Big],
\label{eqn:LQCDffT2}
\end{eqnarray}
where $Q_{\pm}= (M_1 \pm M_2)^2-q^2$.

\subsection{Parametrization of form factors}
\label{sec:2-4}
In our work, we use the form factors obtained using lattice QCD calculations \cite{Meinel:2021mdj, Meinel:2021rbm}. The equations pertinent to our calculations are the nominal form factor, $f (q^2)$, and higher-order form factor, $f_{HO} (q^2)$, which take the following forms in the limit of zero lattice spacing and physical pion mass
\begin{eqnarray}
f(q^2)&=&F^{f}+A^{f}(\omega-1),\nonumber\\
f_{HO}(q^2)&=&F^{f}_{HO}+A^{f}_{HO}(\omega-1),\\ \nonumber
\end{eqnarray}
where the kinematic variable, $w(q^2)=v.v^{\prime}=\dfrac{M_{1}^{2}+M_{2}^{2}-q^{2}}{2M_{1}M_{2}}$. Here, $v$ and $v^\prime$ are the four velocities of the initial and final baryons, respectively. The nominal and higher-order fit parameters describing the $\Lambda_b\to\Lambda_{c}^{*}(2595)$ and $\Lambda_b\to\Lambda_{c}^{*}(2625)$ form factors are given in Table \ref{table:ff}. The unlisted parameters in Table \ref{table:ff} are given by the end-point relations for the form factors in \cite{Meinel:2021mdj}.

\begin{table}[h!]
\caption{Form factor parameters \cite{Meinel:2021mdj} \label{table:ff}}
\begin{ruledtabular}
\begin{tabular}{ c  c  c  c  c }
$f$ & $F^{f}$  & $A^{f}$ & $F^{f}_{HO}$  & $A^{f}_{HO}$\\
\hline
$f_{0}^{(\frac{1}{2}^-)}$ &0.541(48) &-2.18(76) &0.528(58) &-2.10(74) \\
$f_{+}^{(\frac{1}{2}^-)}$ &0.1680(72) &1.09(24) &0.167(11) &1.10(25) \\
$f_{\perp}^{(\frac{1}{2}^-)}$ &  &0.60(18) &  &0.53(19)\\
$g_{0}^{(\frac{1}{2}^-)}$ &0.2207(99) &0.93(25) &0.221(16) &0.86(25)\\
$g_{+}^{(\frac{1}{2}^-)}$ &0.568(48) &-2.31(76) &0.561(58)  &-2.27(76)\\
$g_{\perp}^{(\frac{1}{2}^-)}$ &1.24(13) &-6.7(2.2) &1.22(15) &-6.5(2.2)\\
$h_{+}^{(\frac{1}{2}^-)}$ &0.1889(80) &0.55(21) &0.190(14)  &0.49(22)\\
$h_{\perp}^{(\frac{1}{2}^-)}$ &  &0.91(23) &  &0.89(23)\\
$\tilde{h}_{+}^{(\frac{1}{2}^-)}$ &1.13(13) &-5.8(2.1) &1.12(15)  &-5.8(2.1)\\
$\tilde{h}_{\perp}^{(\frac{1}{2}^-)}$ &0.548(47) &-2.40(81) &0.543(60)  &-2.36(82)\\
$f_{0}^{(\frac{3}{2}^-)}$ &4.20(39) &-25.4(8.0) &4.32(49) &-27.4(8.9)\\
$f_{+}^{(\frac{3}{2}^-)}$ &    &1.26(18) &  &1.23(20)\\
$f_{\perp}^{(\frac{3}{2}^-)}$ &   &2.56(25)  &  &2.58(30)\\
$f_{\perp^{'}}^{(\frac{3}{2}^-)}$ &0.0692(34) &-0.292(80) &0.0701(45) &-0.295(83)\\
$g_{0}^{(\frac{3}{2}^-)}$ &    &1.20(15)  &    &1.21(17)\\
$g_{+}^{(\frac{3}{2}^-)}$ &    &-25.6(6.9)  &   &-26.3(8.1)\\
$g_{\perp}^{(\frac{3}{2}^-)}$ &3.39(33) &-20.2(6.4)  &3.50(39)  &-23.1(7.8)\\
$g_{\perp^{'}}^{(\frac{3}{2}^-)}$ &-0.066(28) &0.65(38)  &-0.065(28)  &0.63(39)\\
$h_{+}^{(\frac{3}{2}^-)}$ &  &2.50(24) &   &2.63(32)\\
$h_{\perp}^{(\frac{1}{2}^-)}$ &  &1.23(15) &  &1.22(18)\\
$h_{\perp^\prime}^{(\frac{1}{2}^-)}$ &-0.02133(95)  &0.036(17) &-0.0220(16)  &0.035(17)\\
$\tilde{h}_{+}^{(\frac{1}{2}^-)}$ &  &-21.7(6.5) &   &-25.0(8.2)\\
$\tilde{h}_{\perp}^{(\frac{1}{2}^-)}$ &3.74(34) &-26.1(7.1) &3.90(42)  &-27.2(8.4)\\
$\tilde{h}_{\perp^\prime}^{(\frac{1}{2}^-)}$ &0.226(30) &-0.48(36) &0.224(35)  &-0.50(37)\\
\end{tabular}
\end{ruledtabular}
\end{table}

\subsection{Helicity Amplitudes}
\label{sec:2-5}
In this section, we present the expressions for the hadronic helicity amplitudes for $B_1 (1/2^+) \to B_{2}^{*} (1/2^-,3/2^-) \ell \bar{\nu}_{\ell}$ decays obtained in terms of form factors and NP couplings, calculated in the rest frame of the parent baryon $B_1$. The helicity amplitudes for the vector, axial-vector, scalar, pseudoscalar, and tensor currents are defined as 
\begin{eqnarray}
H_{\lambda_2,\lambda_W}^{V}&=& (1+C_{V_L}+C_{V_R}) {\epsilon}^{*\mu}(\lambda_W)\left\langle B_2^*  \vert \bar{c}\gamma_{\mu}b \vert B_1 \right\rangle, \nonumber \\
H_{\lambda_2,\lambda_W}^{A}&=& (1+C_{V_L}-C_{V_R}) {\epsilon}^{*\mu}(\lambda_W)\left\langle B_2^*  \vert \bar{c}\gamma_{\mu} \gamma_{5} b \vert B_1 \right\rangle, \nonumber \\
H_{\lambda_2,\lambda_W=0}^{S}&=& (1+C_{S_L}+C_{S_R}) \left\langle B_2^*  \vert \bar{c} b \vert B_1 \right\rangle, \nonumber \\
H_{\lambda_2,\lambda_W=0}^{P}&=& (1+C_{S_L}-C_{S_R}) \left\langle B_2^*  \vert \bar{c} \gamma_{5} b \vert B_1 \right\rangle,\nonumber \\
H_{\lambda_2,\lambda_W,\lambda_{W^\prime}}^{(T)\lambda_1}&=& C_{T}  {\epsilon}^{*\mu}(\lambda_W) {\epsilon}^{*\nu}(\lambda_{W^\prime}) \left\langle B_2^*  \vert \bar{c} i\sigma_{\mu \nu} (1-\gamma_5) b \vert B_1 \right\rangle,
\end{eqnarray}
where $\lambda_{1} (\lambda_{2})$ are the helicities of the parent (daughter) baryon, $\lambda_{W (W^\prime)}$ are the helicities of the $W_{off-shell}$ boson and $\epsilon^\mu$ is the polarization vector of the $W_{off-shell}$. The helicity of the parent baryon is fixed by the relation $\lambda_1=\lambda_2-\lambda_W$.

\subsubsection{${1/2}^{+} \to {1/2}^{-}$}
For the ${1/2}^{+} \to {1/2}^{-}$ transition, we find the relevant helicity amplitudes as 
\begin{eqnarray}
H_{1/2, t}^{V,A}&=& -f_{0}^{(\frac{1}{2}^-) V,A} (1+C_{V_L}\pm C_{V_R}) \dfrac{\sqrt{Q_{\mp}}}{\sqrt{q^2}} (M_1 \pm M_2), \nonumber \\
H_{1/2, 0}^{V,A}&=& -f_{+}^{(\frac{1}{2}^-) V,A} (1+C_{V_L}\pm C_{V_R}) \dfrac{\sqrt{Q_{\pm}}}{\sqrt{q^2}} (M_1 \mp M_2),\nonumber  \\
H_{1/2, 1}^{V,A}&=& f_{\perp}^{(\frac{1}{2}^-) V,A} (1+C_{V_L}\pm C_{V_R}) \sqrt{2Q_{\pm}}~, \nonumber \\
H_{1/2, 0}^{S,P}&=& \mp f_{0}^{(\frac{1}{2}^-) V,A} (C_{S_L} \pm C_{S_R})\dfrac{\sqrt{Q_{\mp}}}{(m_b \mp m_c)} (M_1 \pm M_2) .
\end{eqnarray}
Other helicity amplitudes can be obtained from the relations $H_{-\lambda_{2},-\lambda_{W}}^{V,A}=\mp H_{\lambda_{2},\lambda_{W}}^{V,A}$ and $H_{-\lambda_{2},-\lambda_{W}}^{S,P}=\mp H_{\lambda_{2},\lambda_{W}}^{S,P}$.

The tensor helicity amplitudes are obtained as
\begin{eqnarray}
H^{(T) -1/2}_{-1/2,t,0} &=& -C_{T} \left[ h_{+}^{(\frac{1}{2}^-)} \sqrt{Q_{+}} + \tilde{h}_{+}^{(\frac{1}{2}^-)} \sqrt{Q_{-}} \right], \nonumber \\
H^{(T) +1/2}_{+1/2,t,0} &=& C_{T} \left[ h_{+}^{(\frac{1}{2}^-)} \sqrt{Q_{+}} - \tilde{h}_{+}^{(\frac{1}{2}^-)} \sqrt{Q_{-}} \right], \nonumber \\
H^{(T) -1/2}_{+1/2,t,+1} &=& -C_{T} \frac{\sqrt{2}}{\sqrt{q^2}}         \left[  h_{\perp}^{(\frac{1}{2}^-)} (M_1 - M_2) \sqrt{Q_+} -  \tilde{h}_{\perp}^{(\frac{1}{2}^-)} (M_1 + M_2) \sqrt{Q_-} \right], \nonumber \\
H^{(T) +1/2}_{-1/2,t,-1} &=& C_{T} \frac{\sqrt{2}}{\sqrt{q^2}}         \left[  h_{\perp}^{(\frac{1}{2}^-)} (M_1 - M_2) \sqrt{Q_+} +  \tilde{h}_{\perp}^{(\frac{1}{2}^-)} (M_1 + M_2) \sqrt{Q_-} \right], \nonumber \\
H^{(T) -1/2}_{+1/2,0,+1} &=& C_{T} \frac{\sqrt{2}}{\sqrt{q^2}}         \left[ h_{\perp}^{(\frac{1}{2}^-)} (M_1 - M_2) \sqrt{Q_+} - \tilde{h}_{\perp}^{(\frac{1}{2}^-)} (M_1 + M_2) \sqrt{Q_-}  \right], \nonumber \\
H^{(T) +1/2}_{-1/2,0,-1} &=& C_{T} \frac{\sqrt{2}}{\sqrt{q^2}}         \left[ h_{\perp}^{(\frac{1}{2}^-)} (M_1 - M_2) \sqrt{Q_+} + \tilde{h}_{\perp}^{(\frac{1}{2}^-)} (M_1 + M_2) \sqrt{Q_-}  \right], \nonumber \\
H^{(T) +1/2}_{+1/2,+1,-1} &=& C_{T} \left[ h_{+}^{(\frac{1}{2}^-)} \sqrt{Q_{+}} - \tilde{h}_{+}^{(\frac{1}{2}^-)} \sqrt{Q_{-}} \right], \nonumber \\
H^{(T) -1/2}_{-1/2,+1,-1} &=& -C_{T} \left[ h_{+}^{(\frac{1}{2}^-)} \sqrt{Q_{+}} + \tilde{h}_{+}^{(\frac{1}{2}^-)} \sqrt{Q_{-}} \right].
\end{eqnarray}
The remaining non-zero tensor helicity amplitudes are given by $H^{(T) \lambda_1}_{\lambda_{2}, \lambda_W, \lambda_{W^\prime}}= - H^{(T) \lambda_1}_{\lambda_{2}, \lambda_{W^\prime}, \lambda}$~.

\subsubsection{${1/2}^{+} \to {3/2}^{-}$}
For the ${1/2}^{+} \to {3/2}^{-}$ transition, we obtain the helicity amplitudes as
\begin{eqnarray}
H_{1/2, t}^{V,A}&=& f_{0}^{(\frac{3}{2}^-) V,A} (1+C_{V_L}\pm C_{V_R}) \sqrt{\frac{2}{3} \frac{Q_{\mp}}{q^2}} \frac{(M_1 \mp M_2)}{2}, \nonumber \\
H_{1/2, 0}^{V,A}&=& f_{+}^{(\frac{3}{2}^-) V,A} (1+C_{V_L}\pm C_{V_R}) \sqrt{\frac{2}{3} \frac{Q_{\pm}}{q^2}} \frac{(M_1 \pm M_2)}{2}, \nonumber \\
H_{1/2, 1}^{V,A}&=& -f_{\perp}^{(\frac{3}{2}^-) V,A} (1+C_{V_L}\pm C_{V_R}) \sqrt{\frac{Q_{\pm}}{3}}, \nonumber  \\
H_{3/2, 1}^{V,A}&=& {\pm}f_{\perp^\prime}^{(\frac{3}{2}^-) V,A} (1+C_{V_L}\pm C_{V_R}) \sqrt{Q_{\pm}}, \nonumber  \\
H_{1/2, 0}^{S,P}&=& \pm f_{0}^{(\frac{3}{2}^-) V,A} (C_{S_L}\pm C_{S_R})  \sqrt{\frac{1}{3}} \frac{\sqrt{Q_{\mp}}}{(m_b \mp m_c)}  (M_1 \mp M_2). 
\end{eqnarray}
Other helicity amplitudes can be obtained from the relations $H_{-\lambda_{2},-\lambda_{W}}^{V,A}=\pm H_{\lambda_{2},\lambda_{W}}^{V,A}$ and $H_{-\lambda_{2},-\lambda_{W}}^{S,P}=\pm H_{\lambda_{2},\lambda_{W}}^{S,P}$.

The tensor helicity amplitudes are obtained as 
\begin{eqnarray}
H^{(T) -1/2}_{-1/2,t,0} &=& C_{T} \left[ h_{+}^{(\frac{3}{2}^-)} \sqrt{\frac{Q_{+}}{6}} - \tilde{h}_{+}^{(\frac{3}{2}^-)} \sqrt{\frac{Q_{-}}{6}} \right], \nonumber \\
H^{(T) +1/2}_{+1/2,t,0} &=& C_{T} \left[ h_{+}^{(\frac{3}{2}^-)} \sqrt{\frac{Q_{+}}{6}} + \tilde{h}_{+}^{(\frac{3}{2}^-)} \sqrt{\frac{Q_{-}}{6}} \right], \nonumber \\
H^{(T) -1/2}_{+1/2,t,+1} &=& -C_{T} \left[ h_{\perp}^{(\frac{3}{2}^-)} \sqrt{\frac{Q_{+}}{3}} \frac{(M_1+M_2)}{\sqrt{q^2}} + \tilde{h}_{\perp}^{(\frac{3}{2}^-)} \sqrt{\frac{Q_{-}}{3}} \frac{(M_1-M_2)}{\sqrt{q^2}} \right], \nonumber \\
H^{(T) +1/2}_{-1/2,t,-1} &=& -C_{T} \left[ h_{\perp}^{(\frac{3}{2}^-)} \sqrt{\frac{Q_{+}}{3}} \frac{(M_1+M_2)}{\sqrt{q^2}} - \tilde{h}_{\perp}^{(\frac{3}{2}^-)} \sqrt{\frac{Q_{-}}{3}} \frac{(M_1-M_2)}{\sqrt{q^2}} \right], \nonumber \\
H^{(T) -1/2}_{+1/2,0,+1} &=& -C_{T} \left[ h_{\perp}^{(\frac{3}{2}^-)} \sqrt{\frac{Q_{+}}{3}} \frac{(M_1+M_2)}{\sqrt{q^2}} + \tilde{h}_{\perp}^{(\frac{3}{2}^-)} \sqrt{\frac{Q_{-}}{3}} \frac{(M_1-M_2)}{\sqrt{q^2}} \right], \nonumber \\
H^{(T) +1/2}_{-1/2,0,-1} &=& -C_{T} \left[ h_{+}^{(\frac{3}{2}^-)} \sqrt{\frac{Q_{+}}{6}} + \tilde{h}_{+}^{(\frac{3}{2}^-)} \sqrt{\frac{Q_{-}}{6}} \right], \nonumber \\
H^{(T) +1/2}_{+1/2,+1,-1} &=& -C_{T} \left[ h_{+}^{(\frac{3}{2}^-)} \sqrt{\frac{Q_{+}}{6}} + \tilde{h}_{+}^{(\frac{3}{2}^-)} \sqrt{\frac{Q_{-}}{6}} \nonumber \right], \nonumber\\
H^{(T) -1/2}_{-1/2,+1,-1} &=& -C_{T} \left[ h_{+}^{(\frac{3}{2}^-)} \sqrt{\frac{Q_{+}}{6}} - \tilde{h}_{+}^{(\frac{3}{2}^-)} \sqrt{\frac{Q_{-}}{6}} \nonumber \right], \nonumber\\
H^{(T) +1/2}_{+3/2,t,+1} &=& C_{T} \left[ h_{\perp^\prime}^{(\frac{3}{2}^-)} \sqrt{\frac{Q_{+}}{q^2}} (M_1+M_2) - \tilde{h}_{\perp^\prime}^{(\frac{3}{2}^-)} \sqrt{\frac{Q_{-}}{q^2}} (M_1-M_2) \right], \nonumber  \\
H^{(T) +1/2}_{+3/2,0,+1} &=& C_{T} \left[ h_{\perp^\prime}^{(\frac{3}{2}^-)} \sqrt{\frac{Q_{+}}{q^2}} (M_1+M_2) - \tilde{h}_{\perp^\prime}^{(\frac{3}{2}^-)} \sqrt{\frac{Q_{-}}{q^2}} (M_1-M_2) \right], \nonumber \\
H^{(T) -1/2}_{-3/2,t,-1} &=& C_{T} \left[ h_{\perp^\prime}^{(\frac{3}{2}^-)} \sqrt{\frac{Q_{+}}{q^2}} (M_1+M_2) + \tilde{h}_{\perp^\prime}^{(\frac{3}{2}^-)} \sqrt{\frac{Q_{-}}{q^2}} (M_1-M_2) \right], \nonumber \\
H^{(T) -1/2}_{-3/2,0,-1} &=& -C_{T} \left[ h_{\perp^\prime}^{(\frac{3}{2}^-)} \sqrt{\frac{Q_{+}}{q^2}} (M_1+M_2) + \tilde{h}_{\perp^\prime}^{(\frac{3}{2}^-)} \sqrt{\frac{Q_{-}}{q^2}} (M_1-M_2) \right].
\end{eqnarray}
Similar to the $1/2^+ \to 1/2^-$ transition, the remaining non-zero tensor helicity amplitudes can be obtained using the relation $H^{(T) \lambda_1}_{\lambda_{2}, \lambda_W, \lambda_{W^\prime}}= - H^{(T) \lambda_1}_{\lambda_{2}, \lambda_{W^\prime}, \lambda}$.

The total left-handed helicity amplitudes for the two transitions are given by
\begin{equation}
H_{\lambda_2,\lambda_W} = H_{\lambda_2,\lambda_W}^V -  H_{\lambda_2,\lambda_W}^A,~~~~ 
H_{\lambda_2,0}^{SP}=H_{\lambda_2,0}^S-H_{\lambda_2,0}^P.
\end{equation}

\subsection{Decay distribution and observables}
\label{sec:2-6}
Following the helicity amplitude method outlined in \cite{Shivashankara:2015cta,Datta:2017aue,Li:2016pdv}, the twofold angular distribution for the $B_1\to B_2^{*} \ell \bar{\nu}_\ell$ decay in the presence of NP can be written as 
\begin{eqnarray}
\frac{d^2\Gamma}{dq^2 d \cos \theta_\ell} &=& \frac{G_F^2 |V_{cb}|^2 q^2 \vert \textbf{p}_{B_2^{*}}\vert}{512 \pi^3 m_{B_1}^2} \left( 1-\frac{m_\ell^2}{q^2} \right)^2  \mathcal{A}_{\frac{1}{2}^{+} \to \frac{1}{2}^{-} (\frac{3}{2}^{-})}  ,
\label{eqn:2fold}
\end{eqnarray}
where $\theta_{\ell}$ represents the angle formed by the charged lepton relative to the direction of the off-shell $W$ boson in the dilepton rest frame and $\vert \textbf{p}_{B^{*}_{2}}\vert={\sqrt{Q_{+}Q_{-}}}/{(2 M_1)}$. The quantities $\mathcal{A}_{\frac{1}{2}^{+} \to \frac{1}{2}^{-} (\frac{3}{2}^{-})}$ are given as 
\begin{eqnarray}
	\mathcal{A}_{\frac{1}{2}^{+} \to \frac{1}{2}^{-}}&=& 2 \sin^2 \theta_\ell ( H_{1/2, 0}^2+H_{-1/2, 0}^2  )+ (1-\cos \theta_\ell)^2  H_{1/2, 1}^2  + (1+ \cos \theta_\ell)^2  H_{-1/2, -1}^2   \nonumber \\
	&&+ \frac{m_{\ell}^2}{q^2} \Big[ 2 \cos^2 \theta_\ell ( H_{1/2, 0}^2 + H_{-1/2, 0}^2  ) + \sin^2 \theta_\ell ( H_{1/2, 1}^2 + H_{-1/2, -1}^2  ) \nonumber \\
	&&+ 2 ( H_{1/2, t}^2 + H_{-1/2, t}^2  ) - 4 \cos \theta_\ell ( H_{1/2, t} H_{1/2, 0}+ H_{-1/2, t} H_{-1/2, 0}  ) \Big] \nonumber \\
	&&+ 2 \Big[ ( H_{1/2, 0}^{SP} )^2+ ( H_{-1/2, 0}^{SP}  )^2 \Big] +  \frac{4 m_{\ell}^{2}}{q^2} \Big{\lbrace} 2 \sin^2 \theta_\ell [ ( H^{(T) 1/2}_{1/2,t,0}+H^{(T) 1/2}_{1/2,1,-1} )^2 \nonumber \\
	&&+  ( H^{(T) -1/2}_{-1/2,t,0}+H^{(T) -1/2}_{-1/2,1,-1}  )^2  ] + (1+\cos \theta_\ell)^2 [ ( H^{(T) 1/2}_{-1/2,t,-1}+H^{(T) 1/2}_{-1/2,0,-1} )^2  ]\nonumber\\
	&& + (1-\cos \theta_\ell)^2  [ ( H^{(T) -1/2}_{1/2,1,0}+H^{(T) -1/2}_{1/2,t,1} )^2 ] \Big{\rbrace} + 8 \cos^2 \theta_\ell [ ( H^{(T) 1/2}_{1/2,t,0}+H^{(T) 1/2}_{1/2,1,-1} )^2 \nonumber\\
	&&+ ( H^{(T) -1/2}_{-1/2,t,0}+H^{(T) -1/2}_{-1/2,1,-1}  )^2 ] + 4 \sin^2 \theta_\ell [ ( H^{(T) 1/2}_{-1/2,t,-1}+H^{(T) 1/2}_{-1/2,0,-1} )^2 \nonumber\\
	&&+ ( H^{(T) -1/2}_{1/2,1,0}+H^{(T) -1/2}_{1/2,t,1} )^2  ] + \frac{4m_\ell}{\sqrt{q^2}} \Big[ -\cos \theta_\ell ( H_{1/2, 0} H_{1/2, 0}^{SP} + H_{-1/2, 0}H_{-1/2, 0}^{SP} )\nonumber\\
	&& + ( H_{1/2, t} H_{1/2, 0}^{SP}+ H_{-1/2, t}H_{-1/2, 0}^{SP} ) + {2 \cos^2 \theta_\ell} ( H_{1/2, 0} H^{(T) 1/2}_{1/2,1,-1} + H_{1/2, 0} H^{(T) 1/2}_{1/2,t,0}\nonumber \\
	&& + H_{-1/2, 0} H^{(T) -1/2}_{-1/2,1,-1} + H_{-1/2, 0} H^{(T) -1/2}_{-1/2,t,0}) - {2 \cos \theta_\ell} ( H_{1/2, t} H^{(T) 1/2}_{1/2,1,-1} \nonumber \\
	&&+ H_{1/2, t} H^{(T) 1/2}_{1/2,t,0} + H_{-1/2, t} H^{(T) -1/2}_{-1/2,1,-1} + H_{-1/2, t} H^{(T) -1/2}_{-1/2,t,0}) + {(1-\cos \theta_\ell)^2}\nonumber \\
	&&\times ( H_{1/2, 1} H^{(T) -1/2}_{1/2,1,0} + H_{1/2, 1} H^{(T) -1/2}_{1/2,t,1} ) + {(1+\cos \theta_\ell)^2} ( H_{-1/2, -1} H^{(T) 1/2}_{-1/2,0,-1}\nonumber \\
	&&  + H_{-1/2, -1} H^{(T) 1/2}_{-1/2,t,-1} ) + {\sin^2 \theta_\ell} (H_{1/2, 1} H^{(T) -1/2}_{1/2,1,0} + H_{1/2, 1} H^{(T) -1/2}_{1/2,t,1} + H_{-1/2, -1}\nonumber\\
	&&   \times H^{(T) 1/2}_{-1/2,0,-1}  + H_{-1/2, -1} H^{(T) 1/2}_{-1/2,t,-1} + 2 H_{1/2, 0} H^{(T) 1/2}_{1/2,1,-1} + 2 H_{1/2, 0} H^{(T) 1/2}_{1/2,t,0}\nonumber \\
	&&   + 2 H_{-1/2, 0} H^{(T) -1/2}_{-1/2,1,-1} + 2 H_{-1/2, 0} H^{(T) -1/2}_{-1/2,t,0} ) \Big] -8 \cos \theta_\ell ( H_{1/2, 0}^{SP} H^{(T) 1/2}_{1/2,1,-1}\nonumber \\
	&&    + H_{1/2, 0}^{SP} H^{(T) 1/2}_{1/2,t,0}  + H_{-1/2, 0}^{SP} H^{(T) -1/2}_{-1/2,1,-1} + H_{-1/2, 0}^{SP} H^{(T) -1/2}_{-1/2,t,0} ),
\end{eqnarray}
and
\begin{eqnarray}
	\mathcal{A}_{\frac{1}{2}^{+} \to \frac{3}{2}^{-}}&=& 2 \sin^2 \theta_\ell ( H_{1/2, 0}^2+H_{-1/2, 0}^2  )+ (1-\cos \theta_\ell)^2 ( H_{1/2, 1}^2 + H_{3/2, 1}^2 ) + (1+ \cos \theta_\ell)^2 \nonumber\\
	&&\times  ( H_{-1/2, -1}^2 + H_{-3/2, -1}^2  ) + \frac{m_{\ell}^2}{q^2} \Big[ 2 \cos^2 \theta_\ell ( H_{1/2, 0}^2 + H_{-1/2, 0}^2  ) + \sin^2 \theta_\ell ( H_{1/2, 1}^2 \nonumber \\
	&&  + H_{-1/2, -1}^2 + H_{3/2, 1}^2 + H_{-3/2, -1}^2  ) + 2 ( H_{1/2, t}^2 + H_{-1/2, t}^2  ) - 4 \cos \theta_\ell ( H_{1/2, t} H_{1/2, 0}\nonumber \\
	&& + H_{-1/2, t} H_{-1/2, 0}  ) \Big] + 2 \Big[ ( H_{1/2, 0}^{SP} )^2+ ( H_{-1/2, 0}^{SP}  )^2 \Big] +  \frac{4 m_{\ell}^{2}}{q^2} \Big{\lbrace} 2 \sin^2 \theta_\ell [ ( H^{(T) 1/2}_{1/2,t,0} \nonumber \\
	&&+H^{(T) 1/2}_{1/2,-1,1} )^2 +  ( H^{(T) -1/2}_{-1/2,t,0}+H^{(T) -1/2}_{-1/2,-1,1}  )^2  ] + (1+\cos \theta_\ell)^2 [ ( H^{(T) 1/2}_{-1/2,t,-1} \nonumber\\
	&&+H^{(T) 1/2}_{-1/2,-1,0} )^2 + ( H^{(T) -1/2}_{-3/2,t,-1}+H^{(T) -1/2}_{-3/2,-1,0} )^2  ] + (1-\cos \theta_\ell)^2  [ ( H^{(T) -1/2}_{1/2,1,0} \nonumber \\ 
	&&+H^{(T) -1/2}_{1/2,t,1} )^2 + ( H^{(T) 1/2}_{3/2,0,1}+H^{(T) 1/2}_{3/2,t,1} )^2  ] \Big{\rbrace} + 8 \cos^2 \theta_\ell [ ( H^{(T) 1/2}_{1/2,t,0}+H^{(T) 1/2}_{1/2,-1,1} )^2\nonumber\\
	&& + ( H^{(T) -1/2}_{-1/2,t,0}+H^{(T) -1/2}_{-1/2,-1,1}  )^2 ] + 4 \sin^2 \theta_\ell [ ( H^{(T) 1/2}_{-1/2,t,-1}+H^{(T) 1/2}_{-1/2,-1,0} )^2 + ( H^{(T) -1/2}_{1/2,1,0}\nonumber\\
	&& +H^{(T) -1/2}_{1/2,t,1} )^2  ] + \frac{4m_\ell}{\sqrt{q^2}} \Big[ -\cos \theta_\ell ( H_{1/2, 0} H_{1/2, 0}^{SP} + H_{-1/2, 0}H_{-1/2, 0}^{SP} ) + ( H_{1/2, t} H_{1/2, 0}^{SP}\nonumber\\
	&&+ H_{-1/2, t}H_{-1/2, 0}^{SP} ) + {2 \cos^2 \theta_\ell} ( H_{1/2, 0} H^{(T) 1/2}_{1/2,-1,1} + H_{1/2, 0} H^{(T) 1/2}_{1/2,t,0} + H_{-1/2, 0}\nonumber \\
	&& \times H^{(T) -1/2}_{-1/2,-1,1} + H_{-1/2, 0} H^{(T) -1/2}_{-1/2,t,0}) - {2 \cos \theta_\ell} ( H_{1/2, t} H^{(T) 1/2}_{1/2,-1,1} + H_{1/2, t} H^{(T) 1/2}_{1/2,t,0}\nonumber \\
	&& + H_{-1/2, t} H^{(T) -1/2}_{-1/2,-1,1} + H_{-1/2, t} H^{(T) -1/2}_{-1/2,t,0}) + {(1-\cos \theta_\ell)^2} ( H_{1/2, 1} H^{(T) -1/2}_{1/2,1,0} \nonumber \\
	&& + H_{1/2, 1} H^{(T) -1/2}_{1/2,t,1} + H_{1/2, 1} H^{(T) 1/2}_{3/2,0,1} + H_{1/2, 1} H^{(T) 1/2}_{3/2,t,1}  ) + {(1+\cos \theta_\ell)^2} ( H_{-1/2, -1} \nonumber \\
	&& \times H^{(T) 1/2}_{-1/2,-1,0} + H_{-1/2, -1} H^{(T) 1/2}_{-1/2,t,-1} + H_{-1/2, -1} H^{(T) -1/2}_{-3/2,-1,0} + H_{-1/2, -1} H^{(T) -1/2}_{-3/2,t,-1} )  \nonumber \\
	&& + {\sin^2 \theta_\ell} (H_{1/2, 1} H^{(T) -1/2}_{1/2,1,0} + H_{1/2, 1} H^{(T) -1/2}_{1/2,t,1}  + H_{1/2, 1} H^{(T) 1/2}_{3/2,0,1} + H_{1/2, 1} H^{(T) 1/2}_{3/2,t,1}  \nonumber \\
	&& + H_{-1/2, -1}  H^{(T) 1/2}_{-1/2,-1,0} + H_{-1/2, -1} H^{(T) 1/2}_{-1/2,t,-1} + H_{-1/2, -1} H^{(T) -1/2}_{-3/2,-1,0} + H_{-1/2, -1}\nonumber \\
	&& \times H^{(T) -1/2}_{-3/2,t,-1} + 2 H_{1/2, 0} H^{(T) 1/2}_{1/2,-1,1} + 2 H_{1/2, 0} H^{(T) 1/2}_{1/2,t,0} + 2 H_{-1/2, 0} H^{(T) -1/2}_{-1/2,-1,1} \nonumber \\
	&& + 2 H_{-1/2, 0} H^{(T) -1/2}_{-1/2,t,0} ) \Big]  -8 \cos \theta_\ell ( H_{1/2, 0}^{SP} H^{(T) 1/2}_{1/2,-1,1} + H_{1/2, 0}^{SP} H^{(T) 1/2}_{1/2,t,0} + H_{-1/2, 0}^{SP}  \nonumber\\
	&& \times H^{(T) -1/2}_{-1/2,-1,1} + H_{-1/2, 0}^{SP} H^{(T) -1/2}_{-1/2,t,0} ).
\end{eqnarray} 
The differential decay rate for the $B_1\to B_2^{*} \ell \bar{\nu}_{\ell}$ decay is obtained after integrating out $\cos\theta_\ell$ from Eq. (\ref{eqn:2fold}) as 
\begin{eqnarray}
\frac{d\Gamma}{dq^2} &=& \frac{G_F^2 \vert V_{cb} \vert^2 q^2 \vert \textbf{p}_{B_2^{*}}\vert}{192 \pi^3 m_{B_1}^2} \left( 1-\frac{m_\ell^2}{q^2}\right) ^2  \mathcal{H}_{\frac{1}{2}^{+} \to \frac{1}{2}^{-} (\frac{3}{2}^{-})},
\label{eqn:decayrate}
\end{eqnarray}
where 
\begin{eqnarray}
	\mathcal{H}_{\frac{1}{2}^{+} \to \frac{1}{2}^{-}} &=&   H_{1/2, 0}^2 + H_{-1/2, 0}^2 +  H_{1/2, 1}^2 + H_{-1/2, -1}^2   +\frac{m_\ell^2}{2 q^2} \Big[  H_{1/2, 0}^2 + H_{-1/2, 0}^2 +  H_{1/2, 1}^2   \nonumber \\
	&&+ H_{-1/2, -1}^2 +3( H_{1/2, t}^2 + H_{-1/2, t}^2 ) \Big] + \frac{3}{2} \Big[ ( H_{1/2, 0}^{SP} )^2+ ( H_{-1/2, 0}^{SP}  )^2 \Big]  \nonumber \\
	&& + 2 \left( 1+\frac{2 m_\ell^2}{q^2} \right) \Big[( H^{(T) 1/2}_{1/2,t,0} + H^{(T) 1/2}_{1/2,1,-1} )^2 + ( H^{(T) 1/2}_{-1/2,t,-1}+H^{(T) 1/2}_{-1/2,0,-1} )^2 \nonumber \\
	&& + ( H^{(T) -1/2}_{1/2,1,0}+H^{(T) -1/2}_{1/2,t,1} )^2 +  ( H^{(T) -1/2}_{-1/2,t,0}+H^{(T) -1/2}_{-1/2,1,-1}  )^2 \Big] + \frac{3m_\ell}{\sqrt{q^2}} \nonumber \\
	&& \times \Big[ H_{1/2, 0}^{SP} H_{1/2, t} + H_{-1/2, 0}^{SP} H_{-1/2, t} \Big] + \frac{6m_\ell}{\sqrt{q^2}} \Big[ H_{1/2, 0} ( H^{(T) 1/2}_{1/2,1,-1} + H^{(T) 1/2}_{1/2,t,0} )  \nonumber\\
	&& + H_{-1/2, 0} ( H^{(T) -1/2}_{-1/2,1,-1} + H^{(T) -1/2}_{-1/2,t,0} ) + H_{1/2, 1} ( H^{(T) -1/2}_{1/2,1,0} + H^{(T) -1/2}_{1/2,t,1} ) \nonumber \\
	&& + H_{-1/2, -1} ( H^{(T) 1/2}_{-1/2,0,-1} + H^{(T) 1/2}_{-1/2,t,-1} ) \Big],
\label{TotalHA1}	
\end{eqnarray}
and
\begin{eqnarray}
	\mathcal{H}_{\frac{1}{2}^{+} \to \frac{3}{2}^{-}} &=&   H_{1/2, 0}^2 + H_{-1/2, 0}^2 +  H_{1/2, 1}^2 + H_{-1/2, -1}^2  + H_{3/2, 1}^2 + H_{-3/2, -1}^2  \nonumber \\
	&& +\frac{m_\ell^2}{2 q^2} \Big[  H_{1/2, 0}^2 + H_{-1/2, 0}^2 +  H_{1/2, 1}^2 + H_{-1/2, -1}^2 + H_{3/2, 1}^2 + H_{-3/2, -1}^2  \nonumber \\
	&& +3( H_{1/2, t}^2 + H_{-1/2, t}^2 ) \Big] + \frac{3}{2} \Big[ ( H_{1/2, 0}^{SP} )^2+ ( H_{-1/2, 0}^{SP}  )^2 \Big] + 2 \left( 1+\frac{2 m_\ell^2}{q^2} \right) \nonumber \\
	&& \times  \Big[( H^{(T) 1/2}_{1/2,t,0}+ H^{(T) 1/2}_{1/2,-1,1} )^2 + ( H^{(T) 1/2}_{-1/2,t,-1}+H^{(T) 1/2}_{-1/2,-1,0} )^2 + ( H^{(T) -1/2}_{-3/2,t,-1} \nonumber \\
	&& +H^{(T) -1/2}_{-3/2,-1,0} )^2  
	+ ( H^{(T) -1/2}_{1/2,1,0} + H^{(T) -1/2}_{1/2,t,1} )^2 + ( H^{(T) 1/2}_{3/2,0,1}+H^{(T) 1/2}_{3/2,t,1} )^2   \nonumber \\
	&& +  ( H^{(T) -1/2}_{-1/2,t,0}+H^{(T) -1/2}_{-1/2,-1,1}  )^2 \Big] + \frac{3m_\ell}{\sqrt{q^2}} \Big[ H_{1/2, 0}^{SP} H_{1/2, t} + H_{-1/2, 0}^{SP} H_{-1/2, t} \Big]   \nonumber \\
	&& + \frac{6m_\ell}{\sqrt{q^2}} \Big[ H_{1/2, 0} ( H^{(T) 1/2}_{1/2,-1,1} + H^{(T) 1/2}_{1/2,t,0} ) + H_{-1/2, 0} ( H^{(T) -1/2}_{-1/2,-1,1} + H^{(T) -1/2}_{-1/2,t,0} )  \nonumber \\
	&&  + H_{1/2, 1} ( H^{(T) -1/2}_{1/2,1,0} + H^{(T) -1/2}_{1/2,t,1} +  H^{(T) 1/2}_{3/2,0,1}  +  H^{(T) 1/2}_{3/2,t,1} ) + H_{-1/2, -1} \nonumber \\
	&& \times ( H^{(T) 1/2}_{-1/2,-1,0} + H^{(T) 1/2}_{-1/2,t,-1} +  H^{(T) -1/2}_{-3/2,-1,0} + H^{(T) -1/2}_{-3/2,t,-1} ) \Big].
\label{THA2}	
\end{eqnarray}
Other $q^2$-dependent observables are defined as follows. The ratio of branching fractions is defined as
\begin{equation}
R_{B_2^{*}}(q^2)=\frac{\frac{d\Gamma}{dq^2 }(B_1 \to {B_2^{*}} \tau \bar{\nu}_\tau)}{\frac{d\Gamma}{dq^2 }(B_1 \to {B_2^{*}} \ell \bar{\nu}_{\ell})},
\end{equation}
the forward-backward asymmetry of the charged lepton as
\begin{equation}
A_{FB}^{\tau}(q^2)= \frac{\left( \displaystyle{\int_{0}^{1}}-\displaystyle{\int_{-1}^{0}}\right) \frac{d^2\Gamma}{dq^2 d \cos\theta_{\ell}} d\cos\theta_{\ell}}{\left( \displaystyle{\int_{0}^{1}}+\displaystyle{\int_{-1}^{0}}\right) \frac{d^2\Gamma}{dq^2 d \cos\theta_{\ell}} d\cos\theta_{\ell}},
\end{equation}
the flat term \cite{Boer:2018vpx} as
\begin{equation}
F_{H}^{\tau}(q^2)= \frac{1}{{d\Gamma}/{dq^2}} \displaystyle{\int_{-1}^{+1}} d\cos\theta_{\ell} \left[ \omega_{F_{H}} (\cos\theta_{\ell})\frac{d^2\Gamma}{dq^2 d \cos\theta_\ell} \right]  , 
\end{equation}
where $\omega_{F_{H}} (\cos\theta_{\ell})= 5P_{2} (\cos\theta_{\ell}) + P_{0}(\cos\theta_{\ell})$ are the the weight functions, and $P_{n}$ denote the $n^{\mathrm{th}}$ Legendre polynomial.
Finally, the longitudinal polarization of the charged lepton is given as
\begin{equation}
P_{L}^{\tau}(q^2)=\frac{\frac{d\Gamma^{\lambda_\tau=1/2}}{dq^2}-\frac{d\Gamma^{\lambda_\tau=-1/2}}{dq^2}}{\frac{d\Gamma^{\lambda_\tau=1/2}}{dq^2}+\frac{d\Gamma^{\lambda_\tau=-1/2}}{dq^2}} ,
\end{equation}
where $d\Gamma^{\lambda_\tau= \pm 1/2}/dq^2$ are helicity-dependent differential decay rates.

\section{Fit analysis and $q^2$-spectra}
\label{sec:3}
In this section, we analyze and test the new physics sensitivity of the previously defined $q^2$-dependent observables of $\Lambda_b\to \Lambda_{c}^{*}(2595,2625) \tau^- \bar{\nu}_{\tau}$ decays in the $U_1$ and $S_1$ LQ scenarios. The values of the input parameters used in our analysis are listed in Table \ref{table:inp1}. In calculating the SM estimates, we account for the theoretical uncertainties arising from $V_{cb}$ and associated form factors. Using the nominal fit parameters $F^{f}$ and $A^{f}$ defined in Sec. \ref{sec:2-4}, we compute the central value and statistical uncertainty of the observables. The higher-order fit parameters $F^{f}_{HO}$ and $A^{f}_{HO}$ along with the nominal fit parameters are used to obtain the systematic uncertainties as explained in Eqs. (82)-(84) of \cite{Detmold:2015aaa}. For each observable, we evaluate the statistical and systematic uncertainties associated with the form factors, incorporating the covariance matrices between the form factor parameters as found in \cite{Meinel:2021mdj}. The total uncertainty associated with the form factors is computed by combining these statistical and systematic uncertainties in quadrature.

\begin{table}
\caption{Input parameters\label{table:inp1} \cite{ParticleDataGroup:2024cfk}}
\begin{ruledtabular}
\begin{tabular}{c c c c} 
Parameter &Value &Parameter &Value \\
\hline 
$G_F$ &$1.166378 \times 10^{-5}$ GeV$^{-2}$ &$m_{b}$ &$4.183$ GeV \\
$V_{cb}$ &$(41.1\pm1.2)\times10^{-3}$ &$m_{c}$ &$1.2730$ GeV \\
$V_{tb}$ &$1.010\pm 0.027$ &$m_{\Lambda_b}$ &$5.61960$ GeV \\
$m_{\tau}$ &$1.77693$ GeV &$m_{\Lambda_{c}^{*} (1/2^-)}$ &$2.59225$ GeV \\
$m_{\mu}$ &$0.105658$ GeV &$m_{\Lambda_{c}^{*} (3/2^-)}$ &$2.62800$ GeV\\
$m_{e}$ &$0.51099 \times 10^{-3}$ GeV  & & 
\end{tabular}
\end{ruledtabular}
\end{table}

\begin{table}
\caption{Experimental and theoretical values of $R_{D^{(*)}}, R_{J\psi}, F_{L}^{D^*}, P_{\tau}^{D^*}$\label{table:inp2}}
\begin{ruledtabular}
\begin{tabular}{c c c}
Observable & Experimental value  & SM value \\
\hline
$R_{D}$ &$0.342\pm0.026$ \cite{HFLAV} &$0.298\pm0.004$ \cite{HFLAV} \\
$R_{D^*}$ &$0.287\pm0.012$ \cite{HFLAV} &$0.254\pm0.005$ \cite{HFLAV} \\
$R_{J/\psi}$ &$0.61\pm0.18$ \cite{Iguro:2024hyk}  &$0.258\pm0.004$ \cite{Harrison:2020nrv}\\
$F_{L}^{D^*}$ &$0.49\pm0.05$ \cite{Iguro:2024hyk}  &$0.464\pm0.003$ \cite{Iguro:2020cpg}\\
$P_{\tau}^{D^*}$ &$-0.38\pm0.51^{+0.21}_{-0.16}$ \cite{Belle:2016dyj} &$-0.497\pm0.007$ \cite{Iguro:2020cpg} \\
\end{tabular}
\end{ruledtabular}
\end{table} 

\begin{figure}[h]
\begin{center}
\includegraphics[width=0.45\linewidth]{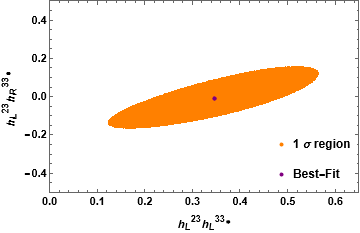}\hfill
\includegraphics[width=0.45\linewidth]{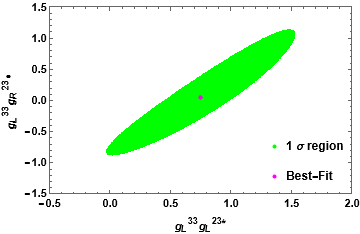}
\caption{$1 \sigma$ constraint region for the $U_1$ (left) and $S_1$ (right) leptoquark couplings at $M_{LQ}=2$ TeV.}
\label{fig:LQ_ps}
\end{center}
\end{figure}

To study the effects of LQs on the decay modes of interest, we first constrain the new couplings by performing a $\chi^2$ analysis using the current experimental measurements of $R_{D^{(*)}}$, $R_{J/ \psi}$, $F_{L}^{D^*}$ and $P_{\tau}^{D^*}$ given in Table \ref{table:inp2}.  
The $\chi^2$ function is defined as
\begin{equation}
\chi^{2}( C_k )=\sum_{ij}^{N_{obs}} [ O_{i}^{exp}-O_{i}^{th}( C_k ) ] \mathcal{V}_{ij}^{-1} [ O_{j}^{exp}-O_{j}^{th}( C_k ) ], \label{chisq}
\end{equation}
where  $O_{i}^{th}$ denotes the theoretical expressions of the observables parametrized in terms of the new couplings $C_k$ as detailed in \cite{Iguro:2024hyk}, while $O_{i}^{exp}$ represents their corresponding experimentally measured values. The correlation of $R_{D}$ and $R_{D^{*}}$ is comprised in the covariance matrix $\mathcal{V}$. Considering an upper bound of $\mathcal{B}(B_{c}^{+}\to \tau^{+}\nu_{\tau})< 30\%$ \cite{Alonso:2016oyd} in the $\chi^2$ fitting, we compute the best-fit values of the combinations of LQ couplings appearing in Eqs. (\ref{CVL_U1})-(\ref{CT_S1}). We take the couplings to be real and the obtained best-fit results are summarized in Table \ref{tab-np}. The $1\sigma$ constraint region permissible for the LQ couplings at $M_{LQ}=2$ TeV is displayed in Fig. \ref{fig:LQ_ps}.
 
\begin{table}[h]
\centering
\caption{Best-fit values of LQ couplings for $M_{LQ}= 2$ TeV. Fitted Wilson coefficients $C_k$ are at the $m_b$ scale.}
\label{tab-np}       
\begin{ruledtabular}
\begin{tabular}{c c c c c c}
 &Best-fit value &Fitted $C_{k}$  &$\chi^{2}_{\textrm{min}}$  &p-value &Pull $(\sigma)$ \\ \hline
$U_1$ &\multirow{2}{*}{\shortstack[c]{$(h_{L}^{23} h_{L}^{33*}, h_{L}^{23}h_{R}^{33*})$\\ $=(0.3473, -0.0116)$}} &\multirow{2}{*}{\shortstack[c]{$C_{V_L}=0.0646$\\$C_{S_R}=0.0086$}} &$3.4194$  &$0.33$  &$3.94$  \\ \\
$S_1$ &\multirow{2}{*}{\shortstack[c]{$(g_{L}^{33} g_{L}^{23*}, g_{L}^{33}g_{R}^{23*})$\\ $=(0.7517,0.0424)$}} &\multirow{2}{*}{\shortstack[c]{$C_{V_L}=0.0699$\\$C_{S_L}(-8.9C_{T})=-0.0077$}} &$3.4271$ &$0.33$ &$3.939$ \\ \\
\end{tabular}
\end{ruledtabular}
\end{table}
 
 To evaluate the goodness-of-fit for each LQ scenario, we compute both the p-value and the pull value. The p-value, as defined in \cite{Blanke:2018yud}, is given by
\begin{equation}
\textrm{p-value}=1- \textrm{CDF}_{n}(\chi^{2}_{\textrm{min}}),
\end{equation}
where $\textrm{CDF}_{n}$ represents the cumulative distribution function of a $\chi^2$ distributed variable with $n$ degrees of freedom, and $\chi^{2}_{\textrm{min}}$ denotes the minimum value of $\chi^2$ at the best-fit point. The p-value statistically assesses the agreement between the theoretical hypothesis and the observed data.
The pull value, expressed in units of standard deviation $\sigma$, is defined as $\sqrt{\chi^{2}_{\textrm{SM}}-\chi^{2}_{\textrm{min}}}$,
where $\chi^{2}_{\textrm{SM}}$ corresponds to the $\chi^2$ value for the Standard Model. This metric measures the statistical enhancement of the fit offered by the LQ scenario relative to the Standard Model. In the SM, we obtain $\chi^{2}_{\textrm{SM}}=18.9434$, corresponding to a deviation of $3.095 \sigma$ and a p-value of $1.97 \times 10^{-3}$. Table \ref{tab-np} shows that both $U_1$ and $S_1$ LQ scenarios yield a p-value of $0.33$, indicating that both models fit the observed data well. 
Defining a quantity
\begin{equation}
d_{O_{i}}=\frac{O_{i}^{LQ}-O_{i}^{exp}}{\sigma_{i}^{exp}} ,
\end{equation}
we also evaluate the discrepancy between the measured data for $R_{D^{(*)}}$, $R_{J/ \psi}$, $F_{L}^{D^*}$, $P_{\tau}^{D^*}$ and their predictions within each LQ model denoted as $O_{i}^{exp}$ and $O_{i}^{LQ}$, respectively. $\sigma_{i}^{exp}$ is the corresponding experimental uncertainty. We present these predictions in Table \ref{tab-pred}.
We find that except for $R_{J/\psi}$, all the other data can be explained within their $1\sigma$ uncertainties in both $U_1$ and $S_1$ LQ scenarios. The $R_{J/\psi}$ data can also be accommodated if the predictions are assessed within $2 \sigma$ confidence interval.

\begin{table}[h]
\centering
\caption{Predictions of observables in the $U_1$ and $S_1$ models using the fitted couplings}
\label{tab-pred}       
\begin{ruledtabular}
\begin{tabular}{c c c c c c}
LQ Scenario  &$R_D$ &$R_{D^*}$ &$R_{J/\psi}$ &$F_{L}^{D^*}$ &$P_{\tau}^{D^*}$ \\ \hline
$U_1$ &\multirow{2}{*}{\shortstack[c]{$0.3418$\\$+0.005\sigma$}} &\multirow{2}{*}{\shortstack[c]{$0.2881$\\$-0.0954\sigma$}} &\multirow{2}{*}{\shortstack[c]{$0.2926$\\$+1.763\sigma$}} &\multirow{2}{*}{\shortstack[c]{$0.4645$\\$+0.5094\sigma$}} &\multirow{2}{*}{\shortstack[c]{$-0.4956$\\$+0.2131\sigma$}} \\ \\
$S_1$ &\multirow{2}{*}{\shortstack[c]{$0.3377$\\$+0.1625\sigma$}} &\multirow{2}{*}{\shortstack[c]{$0.2897$\\$-0.2314\sigma$}} &\multirow{2}{*}{\shortstack[c]{$0.2942$\\$+1.754\sigma$}} &\multirow{2}{*}{\shortstack[c]{$0.4647$\\$+0.504\sigma$}} &\multirow{2}{*}{\shortstack[c]{$-0.4964$\\$+0.2146\sigma$}} \\ \\
\end{tabular}
\end{ruledtabular}
\end{table}

\begin{figure}
\begin{center}
\includegraphics[width=0.3\linewidth]{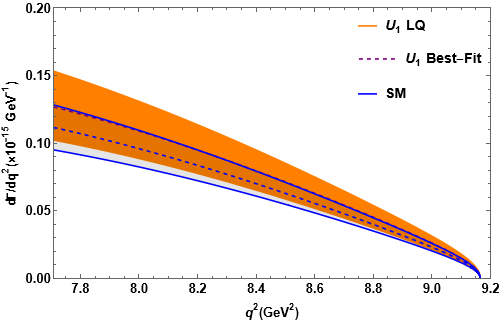}\hfill
\includegraphics[width=0.3\linewidth]{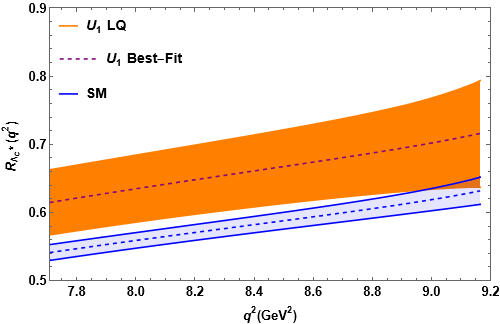}\hfill
\includegraphics[width=0.3\linewidth]{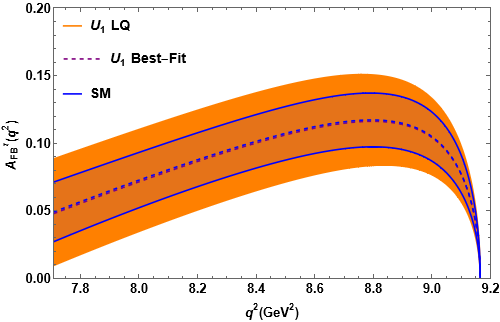}

\vspace{0.5cm}

\includegraphics[width=0.3\linewidth]{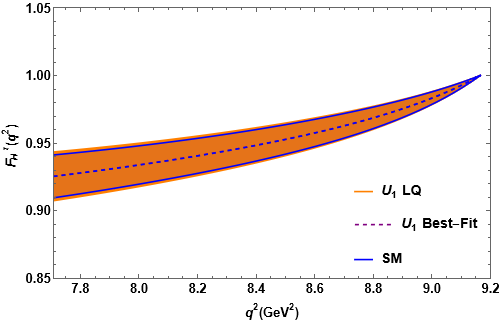}\qquad
\includegraphics[width=0.3\linewidth]{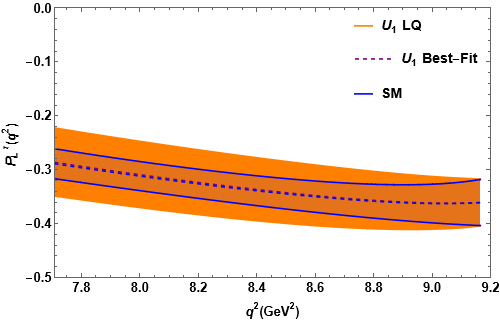}

\vspace{0.5cm}

\includegraphics[width=0.3\linewidth]{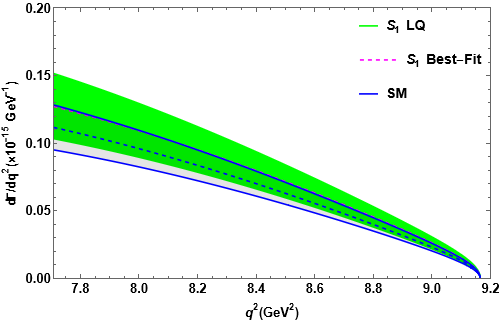}\hfill
\includegraphics[width=0.3\linewidth]{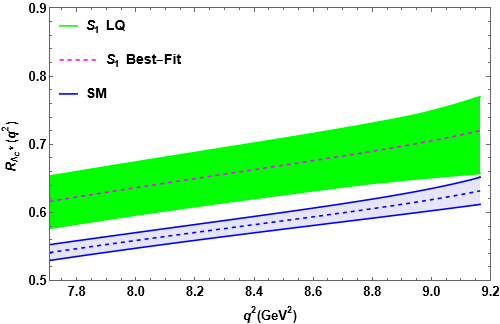}\hfill
\includegraphics[width=0.3\linewidth]{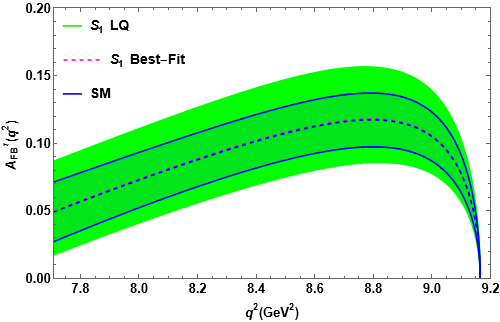}

\vspace{0.5cm}

\includegraphics[width=0.3\linewidth]{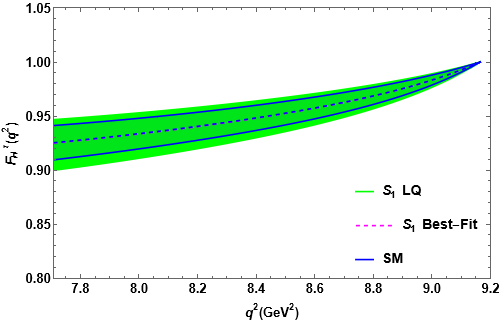}\qquad
\includegraphics[width=0.3\linewidth]{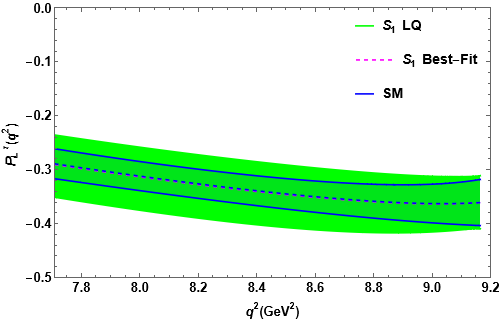} 
\caption{$q^2$-variation of different observables for the $\Lambda_b\to \Lambda_c^{*}(2595) \tau^- \bar{\nu}_{\tau}$ decay mode in the presence of $U_1$ and $S_1$ leptoquarks.}
\label{fig:EL2595}
\end{center}
\end{figure}

\begin{figure}
\begin{center}
\includegraphics[width=0.3\linewidth]{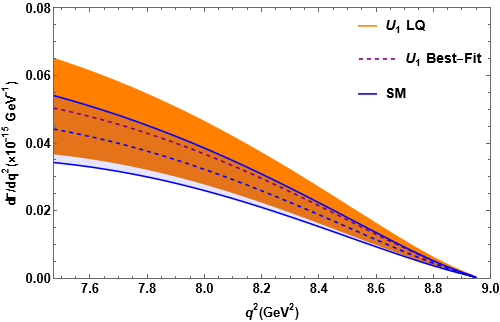}\hfill
\includegraphics[width=0.3\linewidth]{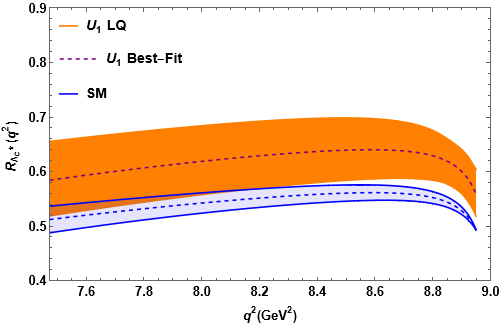}\hfill
\includegraphics[width=0.3\linewidth]{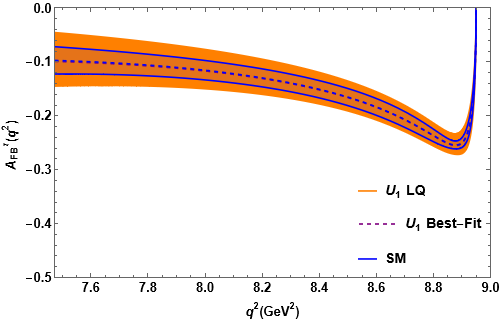}

\vspace{0.5cm}

\includegraphics[width=0.3\linewidth]{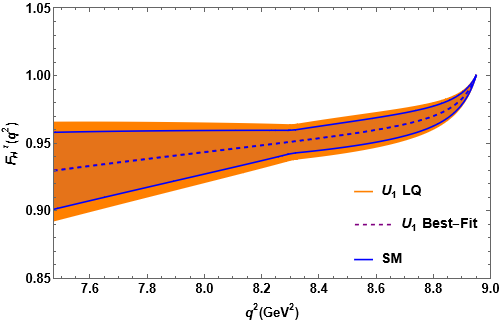}\qquad
\includegraphics[width=0.3\linewidth]{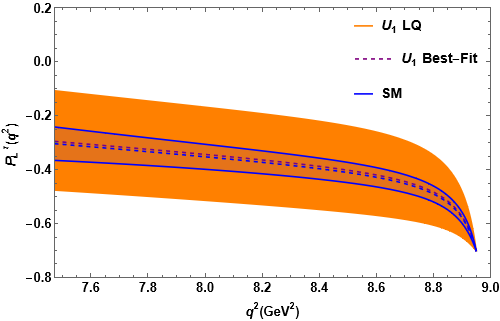}

\vspace{0.5cm}

\includegraphics[width=0.3\linewidth]{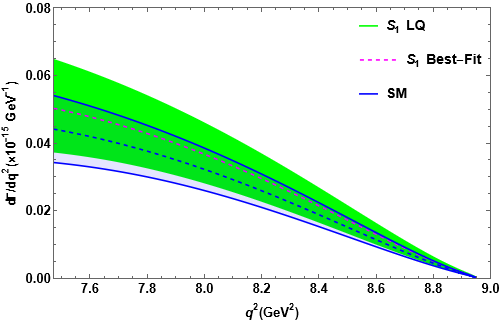}\hfill
\includegraphics[width=0.3\linewidth]{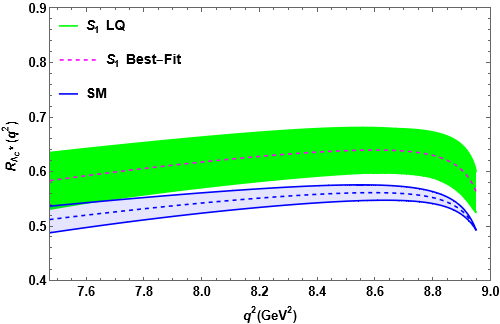}\hfill
\includegraphics[width=0.3\linewidth]{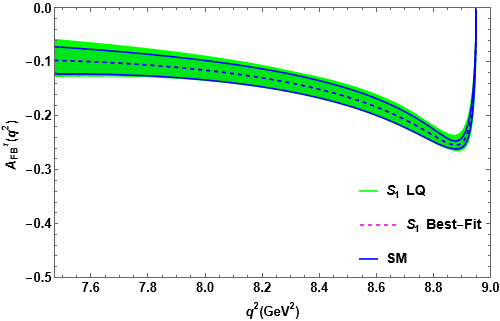}

\vspace{0.5cm}

\includegraphics[width=0.3\linewidth]{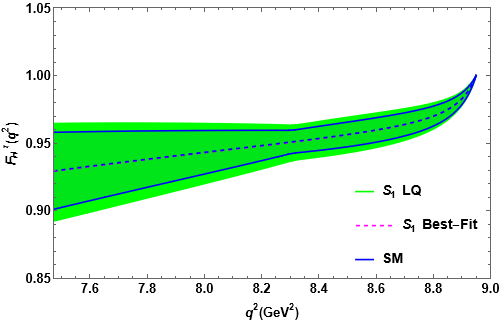}\qquad
\includegraphics[width=0.3\linewidth]{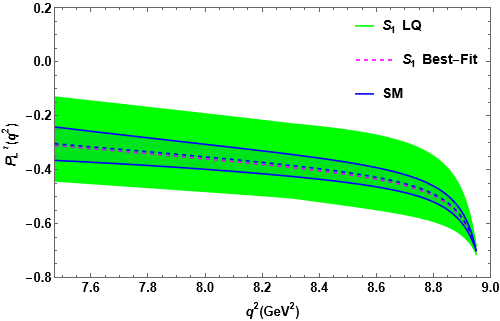} 
\caption{$q^2$-variation of different observables for the $\Lambda_b\to \Lambda_c^{*}(2625) \tau^- \bar{\nu}_{\tau}$ decay mode in the presence of $U_1$ and $S_1$ leptoquarks.}
\label{fig:EL2625}
\end{center}
\end{figure} 

We now present the $q^2$-distributions of various observables for $\Lambda_b\to \Lambda_c^{*}(2595,2625) \tau^- \bar{\nu}_{\tau}$ decays in Figs. \ref{fig:EL2595} and \ref{fig:EL2625}, within the SM and the $U_1$ and $S_1$ LQ scenarios. The LQCD form factors computed in \cite{Meinel:2021mdj} are more reliable in the region near zero recoil. To maintain theoretical robustness and minimize uncertainties, our analysis of the relevant observables is therefore restricted to the high-$q^2$ kinematical region. In Figs. \ref{fig:EL2595} and \ref{fig:EL2625}, the blue bands correspond to the SM predictions, while the orange and green bands represent the effects of the $U_1$ and $S_1$ LQs, respectively. The dashed lines indicate the predictions for the SM central values and for the best-fit values of the new couplings. In both decay channels, the differential decay rate $d\Gamma / dq^2$ is enhanced throughout the considered $q^2$ range in the presence of $U_1$ and $S_1$ LQs. The lepton flavor universality (LFU) ratios $R_{\Lambda_{c}^{*} (2595,2625)}$ show significant deviations from the SM predictions. $R_{\Lambda_{c}^{*} (2595)}$ exhibits a distinguished deviation near the lower end of the $q^2$ spectrum, while $R_{\Lambda_{c}^{*} (2625)}$ shows a similar divergence at higher $q^2$ values. These observations signify the measurement of the LFU ratio in these modes as it can validate the observed anomalies in $b$-decays. Regarding NP sensitivity of the other observables, the best-fit curves of LQs mostly overlap with the SM predictions. However, it can be seen that for $A_{FB}^{\tau}$, the $1\sigma$ band is identifiable from the SM band in $\Lambda_b\to \Lambda_c^{*}(2595) \tau^- \bar{\nu}_{\tau}$, while in the $\Lambda_c^{*}(2625)$ channel, $A_{FB}^{\tau}$ remains largely unaffected by the presence of LQs. The flat term $F_{H}^{\tau}$ displays minimal sensitivity to the LQs for both the $\Lambda_c^{*}(2595)$ and $\Lambda_c^{*}(2625)$ modes. The predictions of lepton polarization $P_{L}^{\tau}$ in both LQ scenarios are distinguishable from the SM band. The restrained NP effects in these observables arise because the factor $\vert 1+C_{V_L} \vert^2$ appearing in both the numerator and denominator largely cancels out. As a result, the NP contributions primarily depend on the scalar couplings, which are small, thereby suppressing any significant deviation. Consequently, the new physics effects in these observables are nearly indistinguishable in the two LQ scenarios.

We also consider the case of complex LQ couplings. The corresponding fitted Wilson coefficients at the $m_b$ scale, obtained for the $U_1$ and $S_1$ LQ models are
\begin{eqnarray}
U_1 &:& C_{V_L} = 0.0576\pm i 0.1225,~C_{S_R}= 0.0086\pm i 0.001,\label{eqn:U1cmplx}\\
S_1 &:& C_{V_L} = 0.0518\pm i 0.3104,~C_{S_L}(=-8.9 C_{T})=-0.1499\pm i 0.6241. \label{eqn:S1cmplx}
\end{eqnarray} 
For the $S_1$ model with complex couplings, the fitted Wilson coefficients of Eq. (\ref{eqn:S1cmplx}) fail to satisfy the LHC bounds provided in \cite{Iguro:2024hyk} and this scenario is thus excluded. However, a restricted case of the $S_1$ model where the couplings are purely imaginary is consistent with these LHC bounds. In this case, the fitted Wilson coefficients at the $m_b$ scale are obtained as
\begin{eqnarray}
S_1 &:& C_{V_L} = i~0.3818,~C_{S_L}(=-8.9 C_{T})= i~0.0261.
\label{eqn:S1pureimg}
\end{eqnarray}
To assess the implications of these complex couplings, we analyze the $q^2$-dependence of the observables in $\Lambda_b\to \Lambda_c^{*}(2595,2625) \tau^- \bar{\nu}_{\tau}$ decays. Using the best-fit values from Eq. (\ref{eqn:U1cmplx}) and (\ref{eqn:S1pureimg}) for the $U_1$ and $S_1$ LQ models, respectively, we find that the predicted behaviour of these observables is very similar to the case with real couplings for the kinematical region considered. We have not shown these results here to avoid redundancy. The observed behaviour can be understood by examining the helicity structure of the decay process. In particular, the leading vector NP contributions enter the decay amplitudes through the factor $\vert 1+C_{V_L} \vert^2$, which is approximately the same for both real and complex couplings. Regarding the scalar couplings, their influence on the decay rate is minimal as their magnitudes are small. The scalar contributions appear in the differential decay rate through the scalar-pseudoscalar helicity amplitudes, $H_{\lambda_2, 0}^{SP}$, as seen from Eq. (\ref{eqn:decayrate}). These terms are further suppressed by a factor of ${m_{\ell}^2}/{q^2}$ or ${m_{\ell}}/{\sqrt{q^2}}$, which becomes more effective near the zero-recoil region. This is the case with real couplings also. Thus, the examined observables are not very sensitive to the scalar couplings near the considered $q^2$ region.

\section{Summary and Conclusion}
\label{sec:4}

With LQs as favorable candidates to address the tensions present in $B \to D^{(*)} \tau \bar{\nu}_{\tau}$ decays, we were inspired to investigate their effects on decays with similar $b \to c \ell^- \bar{\nu}_{\ell}$ quark-level transition. In this work, we analyzed the $\Lambda_b\to \Lambda_{c}^{*}(2595,2625) \tau^- \bar{\nu}_{\tau}$ decay modes within the the $U_1$ and $S_1$ leptoquark frameworks. From the interaction Lagrangian of these LQs with the SM fermions, we obtained the Wilson coefficients contributing to the $b \to c \tau \bar{\nu}_\tau$ transition. We performed a comprehensive analysis of the vector, axial-vector, scalar, pseudo-scalar and tensor currents for the $1/2^+ \to 1/2^-$ and $1/2^+ \to 3/2^-$ transitions of $\Lambda_b\to \Lambda_{c}^{*} \ell^- \bar{\nu}_{\ell}$, and presented the corresponding helicity amplitudes in this paper. We obtained the expressions for the differential decay rate and related observables in the presence of NP. The observables were then analyzed within the SM and the $U_1$ and $S_1$ LQ scenarios. In the numerical analysis, we used a $1\sigma$ bound from the current experimental data of $R_{D^{(*)}}, R_{J/ \psi}, F_{L}^{D^*}, P_{\tau}^{D^*}$ to obtain the LQ parameter space and we have fitted the LQ couplings using a $\chi^2$ method. We found that the current data can be explained with both LQ scenarios within $1 \sigma$ using the best-fit values, except for $R_{J/\psi}$. For the $\Lambda_b\to \Lambda_{c}^{*}(2595,2625) \tau^- \bar{\nu}_{\tau}$ modes, we employed form factors obtained in the LQCD framework to predict the $q^2$-variation of observables like the differential decay rate, the ratio of branching fractions, forward-backward asymmetry, flat term, and longitudinal polarization of the charged lepton. We found that while the angular observables exhibit only mild sensitivity to the two LQs, the LFU ratio shows significant sensitivity to the presence of these LQs. A distinguished divergence from the SM prediction was identified for the LFU ratio within the considered kinematical region for both decay modes. Thus, measuring the LFU ratio $R_{\Lambda_{c}^{*}}$ can shed further light on LFU violation and provide complimentary probe of new physics in the $b \to c$ sector.  Although it may be difficult to distinguish between the $U_1$ and $S_1$ LQs using the $\Lambda_b\to \Lambda_c^{*} \tau^- \bar{\nu}_{\tau}$ decays, with the copious production of $\Lambda_b$ baryons in LHCb, these modes can still be investigated for the presence of LQs.

\begin{acknowledgments}
C P Haritha acknowledges UGC, Govt. of India for the support through fellowship No. F.82-44/2020 (SA -III). B. Mawlong acknowledges the support of the Anusandhan National Research Foundation (ANRF), Govt. of India through research grant no. ANRF/IRG/2024/000256/PS.
\end{acknowledgments}

\bibliography{Lambda_ref}

\end{document}